\documentclass{article}
\usepackage{amssymb}

\begin{document}
\newtheorem{defi}{Definition}
\newtheorem{lem}{Lemma}

\title{\bf Reduction of a Quantum Theory}
\author{Tim A. Koslowski\\{\texttt{tim@physik.uni-wuerzburg.de}}\\Institut f\"ur Theoretische Physik und Astrophysik\\
Universit\"at W\"urzburg\\ Am Hubland\\ 97074 W\"urzburg}
\date{\today}
\maketitle\thispagestyle{empty}

\begin{abstract}
This paper serves as a preparation of work that focuses on extracting cosmological sectors from Loop Quantum Gravity. We start with studying the extraction of subsystems from classical systems. A classical Hamiltonian system can be reduced to a subsystem of ''relevant observables'' using the pull-back under the Poisson-embedding of the ''relevant part of phase space'' into full phase space. Since a quantum theory can be thought of as a noncommutative phase space, one encounters the problem of embedding noncommutative spaces. We solve this problem for a physically interesting set of quantum systems and embeddings by constructing the noncommutative analogue of the construction of an embedding as the projection to the base space of an embedding of fibre bundles over the involved spaces. This paper focuses on the physical ideas that enter our programme of reduction of quantum theories and tries to explain these on examples rather than abstractly, which will be the focus of a forthcoming paper.
\end{abstract}

\section{Introduction}

Although there are viable and well understood theoretical models for real world physical systems, these are usually too complex to answer even very simple questions such as whether a certain chemical substance is a poison. In this case, although the underlying model, quantum mechanics and classical electrodynamics, is very well understood by theoretical physicists, it is better to ask a biochemist about a poison than a theoretical physicist. The abstract reason behind this is that a biochemist will have the more adapted theory at his disposal, whereas the theoretical physicist, although he has the ''more fundamental'' theory at his disposal, is simply overwhelmed by the complexity of the question. He will however probably play an important role in the ''explanation'' of the biochemists theory in therms of a ''fundamental theory''. 

This example illustrates the very nature of ''fundamental theories'' in the sense that although once a ''relevant sector'' or ''dominant contribution'' is identified, it is practically impossible to describe many phenomena exactly in the full theory. On the other hand one can often use the fundamental theory to show how the relevant sector comes about. Ultimately one is drawn to the conclusion, that the explanation of many real world problems becomes further and further removed from the theory the more fundamental the theory becomes. Moreover, as the systems under consideration and particularly the number of degrees of freedom increase from finitely many in a mechanical system to infinitely many in a field theory, one is left with no other choice than to extract relevant degrees of freedom. Already in an interacting field theory, one needs to extract the ''relevant sector'' and try to calculate stable predictions for it, as it is usually done by decomposing the system into clusters of particles that do not interact with the outside world.

This reduction process becomes even more important in a theory about the entire universe, particularly gravity. Here there is no obvious cluster with a small number of particles or deviations thereof, but the theory describes the entire universe at once, even the calculation that is preformed in it, which will pose obvious self descriptional problems in the sense of G\"odel. Thus, it is particularly important in a ''fundamental theory of gravity'' to be able to extract the relevant sector. This is, where this line of investigation started. The particular starting point was Loop Quantum Gravity and the feeling, that one needs to find the cosmological sector of the theory i.e. focusing on the homogeneous isotropic component of space time in this particular framework.

The extraction of a ''relevant sector'' from a ''fundamental'' quantum theory is the goal if this paper. The idea behind this is the following: Given a classical system, we have a phase space, a Poisson-bracket and certain functions on phase space, that represent the generators for certain transformations. Now let us assume, that we have a certain ''relevant sector'' identified by a set of measurable quantities that we can resolve with a given setup. These correspond to a certain set of observables in the Poisson system. Thus, that a sector is ''relevant'' means here that our measurements are ''sensitive to'' it. We can then investigate the subsystem of observables in our Poisson system, that we are sensitive to. Let us furthermore assume, that this subsystem is closed under the Poisson bracket, then we are able to view the system, that we are sensitive to as a Poisson subsystem. Clearly, all the Poisson brackets calculated in the subsystem will coincide with the ones in the full system. Let us now put this construction on its head and consider a reduced system, that for some reason catches the ''relevant'' degrees of freedom, then it is legitimate to ask when it is a Poisson closed subsystem of a more ''fundamental'' system. Obviously, we are looking for a Poisson embedding of the phase space of our reduced ''relevant'' model (or the ''observables that our measurements are sensitive to'') into the more fundamental theory.

On the other hand given a Poisson embedding, that catches the relevant degrees of freedom, then we are able to use the pull-back under the Poisson-embedding to reduce the fundamental system to the relevant system. This observation is the point of departure in this paper: We will consider a certain class of Poison systems and certain Poisson-embeddings of ''smaller'' into ''larger'' Poisson systems of this kind and we will consider associated quantum systems. These will be modelled by a certain class of $C^*$-algebras together with their Hilbert-space representations. By our choice, we will be able to associate the Poisson-systems in a one to one way with $C^*$-algebras, however since we include cases, in which the Stone-von-Neumann theorem does not hold, the Hilbert space representation of these will not be unique. Thus, when we want to lift the concept of Poisson-embedding, we do not only have to consider how to construct the reduced $C^*$-algebra, but we also have to be able to calculate its induced Hilbert space representation.

The particular class of systems that we want to consider are locally compact Lie groups whose generators act as smooth momentum vector fields on a locally compact configuration manifold. In this case, the Poisson system is dual to the transformation Lie algebroid. These structures can be quantized uniquely and contain a large number of physically interesting examples. Moreover, we will be able, by passing to the associated Lie groupoids, to refine our view of ''interesting'' Poisson embeddings. For this class of embeddings we will then use techniques that are related to Morita-equivalence structures to construct the reduced $C^*$-algebra and the induced Hilbert space representations.

The focus of this paper is on the physical ideas which we explain on examples, which demonstrate instances of a more general mathematical setting, that we want to introduce. The focus of this work is not on presenting the mathematical framework, which we will do in a forthcoming paper. For this purpose, we will often switch between physical terms, such as ''quantum observable algebra'' and the mathematical categories in this case ''$C^*$-algebras''. Similarly, we will refer to a Hermitian structure on a $C^*$-module as rank-one operators, which is the way in which we introduce the physical concept, even when these are not operators of rank one. We hope that these deviations from mathematical rigor help to achieve the higher goal of making the physical ideas more transparent.

This paper is organized as follows: Section 2 serves as a revision of the classical setting, where we develop the idea of using the pull-back under a Poisson embedding to construct a reduced system, which we interpret physically as reduced sensitivity of the measurements at our disposal. We collect the requisites that we impose for our quantization of a Poisson embedding in section 3. Section 4 gives an example for Morita theory for $C^*$-algebras using ordinary quantum mechanics form which we extract some mathematical framework that we use to state our strategy more technically. In section 5 we use Lie-transformation groupoids as linking structures, i.e. we use them to mediate between classical and quantum theories, which allows us to define a precise notion of classical limit. Section 6 uses Morita theory for groupoids to develop the main idea of the paper, which is introduced in section 7 again by an example from ordinary quantum mechanics form which we develop the general idea and show its basic properties. We conclude in section 8 with a brief discussion of the imposition of constraints.

\section{Extraction of Relevant Degrees of Freedom in a Classical Theory}

The aim of this paper is to extract ''relevant degrees of freedom from a quantum theory''. Before we specify what exactly we mean, we we will use the often helpful trick to consider the associated classical problem. For quantization it is particularly useful to consider the analogue problem in a Poisson system and use the close relation between the Poisson algebra of classical variables and its quantum counterpart. Therefore, we will first consider a classical canonical Poisson system form which we want to extract a set of ''relevant degrees of freedom''. 

\subsection{Classical System}

In this paper we will consider certain Poisson systems on an algebra of functions on a phase space as a model for classical systems, which we want to ''reduce to a relevant subsystem''. Before defining phase space reduction, let us define a Poisson-structure on functions on a phase space $\Gamma$, which we assume to be a smooth manifold (an exposition of Poisson manifolds can be found in \cite{cecile}). A Poison structure on $\Gamma$ is then a bilinear, antisymmetric map $\{.,.\}$ from $C^\infty(\Gamma)\times C^\infty(\Gamma) \rightarrow C^\infty(\Gamma)$, which satisfies the Jacobi identity and defines for each $f \in C^\infty(\Gamma)$ a vector-field on $\Gamma$ by $\{f,.\}$. These vector-fields are called Hamilton-vector-fields of the respective $f$.
\\
Let $(\Gamma_1,\{.,.\}_1)$ and $(\Gamma_2,\{.,.\}_2)$ be two Poisson manifolds. Let furthermore $i$ be an immersion of $\Gamma_1$ into $\Gamma_2$. We say, that $\Gamma_1$ is a Poisson-submanifold of $\Gamma_2$ if the inclusion map satisfies:
\begin{equation}
  \{f\circ i, g \circ i\}_2 = \{f,g\}_1 \circ i.
\end{equation}
If $(\Gamma_1,\{.,.\}_1)$ is a Poisson-submanifold of $(\Gamma_2,\{.,.\}_2)$, then there are two important consequences: First, the Poisson-structure $\{.,.\}_1$ is unique and second, the Hamilton-vector-fields on $\Gamma_1$ coincide with their push-forward on $\Gamma_2$. The physical interpretation of these two statements is, that first the canonical system on the reduced phase space $\Gamma_1$ is uniquely determined by the canonical structure on the full phase space $\Gamma_2$ and second, that the kinematics of the observables of the reduced system {\bf does coincide} with the kinematics of the corresponding observables of the full system. 
\begin{defi}
  Classical phase space reduction of a full $(\Gamma_2,\{.,.\}_2)$ to $(\Gamma_1,\{.,.\}_1)$ is the pull-back of observables from $\Gamma_2$ under a Poisson-inclusion map.
\end{defi}
An important example of phase spaces is given by symplectic systems on cotangent bundles over a configuration manifold together with its canonical Poisson-structure, stemming from the canonical symplectic form. There, we can easily split the data $(T^*(\mathbb X),\{.,.\})$ into a configuration space $\mathbb X$ and a Lie-algebra of vector fields on $\mathbb X$, which arises as $\{f,.\}$, where $f$ is linear in the fibre. The important point is, that the functions on the configuration space together with the functions linear in the fibre separate the points in phase space. 
\\
We can do a similar decomposition for any Poisson-system $(\Gamma,\{.,.\})$, because we are able to find a maximal Poisson-commuting subspace in each phase space, which we call configuration space. Note, that the choice of configuration space $\mathbb X$ is not uniquely determined by $(\Gamma,\{.,.\})$. Moreover, we will find a set of vector fields on $\mathbb X$ of the kind $\{f,.\}$, for a set of functions, that separates the points in all of $\Gamma$. For technical reasons, we will need a locally compact configuration space. These requirements are our definition of a strong Poisson system, which we will later replace by the more general concept of Lie-algebroids:
\begin{defi}
  Given a smooth, locally compact manifold $\mathbb X$, together with a Lie-algebra $\mathcal L$ of vector fields on $\mathbb X$ is called {\bf a strong Poisson system}.
\end{defi}
Our first goal is to reduce these classical systems so we see the structure behind the reduction process, which we want to generalize to quantum systems thereafter.

\subsection{Extraction of Relevant Degrees of Freedom using Poisson embeddings}

Given two strong symplectic systems $(\mathbb X_1,\mathcal L_1)$and $(\mathbb X_2,\mathcal L_2)$, we define a strong Poisson map in complete analogy to a restriction of a general Poisson system as an immersion $i$ from $(\mathbb X_1,\mathcal L_1)$ to $(\mathbb X_2,\mathcal L_2)$, such that the push-forward of the vector fields (acting on $\mathbb X_i$) coincides with the image of the vector field, such that the Lie-algebra $\mathcal L_1$ is embedded into $\mathcal L_2$ with equivalent action on the image of $i(\mathbb X_1)$.
\\
Let us define two strong Poisson-systems to be isomorphic, iff there exist two Poisson-immersions going in opposite ways between the two systems. This allows us to define isomorphism classes of Poisson systems $[.]_\sim$.
\\
Fixing a particular representative (using the axiom of choice) in each isomorphism class, we see, that the identity embedding of $[(\mathbb X,\mathcal L)]_\sim$ into itself is a strong Poisson map in the obvious way. Moreover, it is obvious, by construction, that the composition of two strong Poisson maps yields another strong Poisson map (up to a possible strong Poisson-isomorphism). This lets us conclude:
\begin{lem}
  Taking isomorphism classes of strong Poisson systems $(\mathbb X,\mathcal L)$ as objects and strong Poisson immersions $i$ as morphisms, we have a well defined category of strong Poisson systems.
\end{lem}
By defining $[1]_\sim \le [2]_\sim$, we notice, that the category of strong Poisson systems is partially ordered. Since the Poisson-immersions are exactly the structure, that allows us to calculate certain observables of a ''larger'' systems in the framework of the ''smaller'' immersed systems, we see that the underlying embedding used to define this partial ordering is exactly what we desired in the introduction.
\\
Our particular goal is to construct these embedded subsystems using a Poisson embedding. For this purpose and for later analogy to the construction in a quantum system, we will apply this construction directly to the algebra of functions on phase space, whose real elements represent the observables of the system.

\subsection{Reduced Sensitivity}

The great advantage of working with Poisson systems is, that they are already formulated in terms of an observable algebra, such that we could forget about the underlying phase space. Obviously, since we have a commutative algebra, we could always look for an enveloping $C^*$-algebra and recover the phase space by the Gelfand-Naimark theorem.
\\
Let us now consider the implications of working with an algebra of observables: Given an observable $f$ on the ''full system'' we can easily construct an observable on the ''reduced system'' by considering its pull-back under the Poisson embedding $\eta$, this is nothing else than to say that $\eta^* f$ is an observable given $f$ is an observable. We can apply this to the entire observable algebra $\mathfrak A$ on the full system and this procedure yields the entire reduced observable algebra $\mathfrak A_{red}$:
\begin{equation}
  \mathfrak A = \{ \eta^* f : f \in \mathfrak A \}.
\end{equation}
This is a well defined quotient of the full algebra by the ideal of functions that vanish at the embedding and hence inherits the entire algebraic structure.
\\
{\bf Hence:} The pull-back under a Poisson embedding is equivalent with the construction of an embeddable reduced Poisson system, whose Hamilton vector fields coincide with the ones of the full system at the embedding. 
\\
In other words: the pull-back under a Poisson-embedding is exactly the construction, that we are seeking.
\\
Let us briefly consider the physical implications of this way of considering the reduction problem: The physical interpretation of an observable is to correspond to the value of a measurement on the system, i.e. given the state of the system is given by a distribution $\mathfrak d$ on phase space, then the expectation value $\mathfrak d(f)=\int_{\Gamma} \mu_{\mathfrak d} f$ represents the expectation value of the outcome of our measurement. The pull-back under a Poisson embedding is then the restriction of the sensitivity of our measurement. Turning this statement around, we can replace the application of ''reducing a given system to a relevant part'' or ''extracting the relevant degrees of freedom'' by ''assuming a reduced sensitivity of the measurements at our disposal''. 
\\
Mathematically, we say that our reduced observables are those that are insensitive to the ideal of observables, that vanish at the embedded phase space. Using this point of view lets us generalize the reduction problem to quantum systems.

\section{Statement of the Problem: Quantize Poisson Embeddings}

The goal of this paper is the extension of ''phase space reduction'' to quantum systems. The general strategy to ''quantize a statement'' is to formulate the statement for a Poisson system, i.e. stating it as a problem on the algebra of observables and to reformulate this statement in such a way that the commutativity of the observable algebra becomes irrelevant.
\\
With the preparations in the previous section, we have already stated the problem in terms of the observable algebra and we can say that we want to construct the quantum analogue of the pull-back under a Poisson embedding. The problem with this statement is however that the ''phase space of quantum mechanics is noncommutative''\footnote{We will make extensive use of some basic ideas of noncommutative geometry, as they are explained e.g. in \cite{connes,varilly1997,landi}}, i.e. it is not a topological space but rather a noncommutative algebra of observables, which is thought of as the algebra of continuous functions on a noncommutative phase space.
\\
If we primitively apply the Gelfand Naimark theorem we proceed as follows: A commutative $C^*$-algebra is isomorphic to the algebra of continuous functions on its spectrum, which is a locally compact Hausdorff space in the Gelfand topology. Thus we could conclude that we are looking for the pullback under an embedding of spectra. This procedure is however flawed as we can see by considering a rather simple example: Take the $C^*$-algebra of ordinary quantum mechanics on any locally compact group, then by the Stone-von Neumann theorem we know, that there is only one unitary equivalence class of regular irreducible representations of this algebra, thus the spectrum consists of only one point. This means that this quantum system is embeddable into any other $C^*$-algebra, particularly into $\mathbb C$, which we view as a degenerate quantum system, whose phase space consists of one point only. Thus $\mathbb D_2$, the algebra of diagonal $2\times 2$-matrices, is ''larger'' than any Heisenberg system. This is in clear contradiction to the idea that embeddability defines a partial ordering.
\\
We have a much better understanding using the reduced sensitivity point of view: We are able to define an ideal of observables that our measurements are insensitive to such that the observable algebra of the reduced quantum system arises as the quotient of the full observable algebra by this ideal. This would be able to define the observable algebra, however other than in a classical theory, where all pure states are evaluations at points in phase space, we need a Hilbert-space representation for the observable algebra in addition to the algebra itself, whose elements represent the pure states of the system. Furthermore, we know that there are in general many inequivalent representations of a given $C^*$-algebra. This shows that we can not separate the reduction of the observable algebra from the reduction of its Hilbert space representation.
\\
A strategy that we could follow to construct a reduced quantum algebra and its Hilbert-space representation is to ''try to read the rules of quantization off'', the to reduce the classical theory using the pull-back under a Poisson embedding and then quantizing this system ''using the extracted rules for quantization''. There are however serious problems with the ''reading off'' of the ''rules of quantization'' since there are many different ways of constructing a $C^*$-algebra and its Hilbert-space representation from a given classical system, and it is in no way clear that using two sets of rules that yield the same full quantum system yield equivalent quantum systems for a reduced system.
\\
Consequently, we do not want to ''reduce and quantize again'', but we want to reduce a quantum system, i.e. given a triple $(\mathfrak A,\pi,\mathcal H)$, where $\mathfrak A$ is a $C^*$-algebra representing the quantum observables of a system and $\pi$ is a representation of this algebra as a subset of the bounded operators on a Hilbert-space $\mathcal H$, we seek the construction $\mathcal E$ of the reduced system $(\mathfrak A_o,\pi_o,\mathcal H_o)$ directly form $(\mathfrak A,\pi,\mathcal H)$\footnote{For an introduction to the application of $C^*$-algebras and their representations in physics see \cite{bratteli1,bratteli2}.}. 
\\
Let us now formalize the requirements of the quantum Poisson map $\mathcal E_\eta$, corresponding to a classical map $\eta$:
\\
First, we want that the construction reproduces the right classical limit, i.e. we want for the observable algebras that the following diagram commutes:
\begin{equation}
    \label{l1}
    \begin{array}{rcccl}
      &&\mathcal E && \\
      & \mathfrak A & \longrightarrow & \mathfrak A_o& \\
      \hbar \rightarrow 0& \downarrow & \eta^* & \downarrow & \hbar \rightarrow 0 \\
      & C_c^\infty(\Gamma) & \longrightarrow & C_c^\infty(\Gamma_o). &
    \end{array}
\end{equation}
Here $\Gamma,\Gamma_o$ are the full and reduced phase space respectively, $\eta$ denotes the embedding of the reduced into the full phase space and $\mathfrak A,\mathfrak A_o$ are the $C^*$-algebras that represent the quantum observables of the associated quantum systems. This diagram needs some explanation: We need to specify the way in which we take the classical limit indicated by $\hbar \rightarrow 0$. Our notion of classical limit will be fixed when we restrict ourselves to transformation group systems, i.e. classical systems in which we give a polarization, such that we are able to talk about a configuration space and second by considering the group generated by the exponentiated Poisson action of the momenta. There is a simple correspondence between the classical Poisson systems and the associated $C^*$-algebras, such that taking the classical limit is an easy procedure. 
\\
Second, we want to construct the ''right Hilbert space representation'': Physically we want that the expectation values of our observables are matched by corresponding expectation values in the full theory. We can reduce the number of assumptions by noticing that any representation of a $C^*$-algebra arises as a direct sum of cyclic representations out of vacuum states $\Omega_i$, where the summands are of the kind $\langle \psi_a, \pi(b) \psi_c \rangle_i=\Omega_i(a^*bc)$. This allows us to restrict our attention to vacuum expectation values, thus we demand that there is a dense subset $\mathcal D$ in the reduced $C^*$-algebra and that there is a vacuum state $\omega_i$, corresponding to $\Omega_i$, on the reduced quantum algebra such that a map $\mathcal E: \mathcal D\subset \mathfrak A_o \rightarrow \mathfrak A$ matches vacuum expectation values:
\begin{equation}
  \label{l2}
  \omega_i(a)=\Omega_i(\mathcal E(a)) \,\,\, \forall a \in \mathcal D.
\end{equation}
This condition ensures that expectation values of the reduced Hilbert space representation is a subset of the expectation values in the full theory, as we would expect it from being a subsystem.
\\
Third, one would like to constrain the dynamics to coincide with the dynamics of the full theory. Let us consider the corresponding classical situation: Given a Poisson submanifold embedded into a larger Poisson manifold, it is generally not the case that the Hamilton vector field of the full Hamiltonian will be tangential to the submanifold. The situation for quantum theories is analogous: Consider the von-Neumann equation for a density operator $\rho$, which reads using the correspondence map $\mathcal E$:
\begin{equation}
  i \partial_t \mathcal E(\rho) = [H,\mathcal E(\rho)],
\end{equation}
which implies that if $H$ contains ''mixing matrix elements'' one obtains a dynamics that moves away from the image of $\mathcal E$. This forces us to use the reduced sensitivity interpretation, which tells us that our measurements are insensitive to an ideal of observables, and that our dynamics has to be corrected by building the quotient. Since the ultimate goal of this work is the extraction of a sector from Quantum Gravity, which is a theory with constrained dynamics, we will not discuss details about dynamics but rather focus on the imposition of constraints (see section 8).
\\
Our strategy to construct a quantum Poisson embedding will be as follows: We will restrict our attention to Lie-transformation group systems and use groupoids as the lining structure between classical and quantum systems. Lie-transformation groupoids are very useful for this purpose, because they are on the one hand classical spaces, which we are able to treat with methods of topology, but on the other hand, one can define a noncommutative algebra of functions on a groupoid using the convolution product, which is precisely the product of the corresponding quantum algebra. Another feature of groupoids is that they act on spaces in a way very similar to the representation of a $C^*$-algebra on a Hilbert-space. This allows one to use Morita theory, i.e. the theory of categories of isomorphism classes of representations of groupoids. It turns out that Morita theory for groupoids \cite{muhly} is very similar to Morita theory for the corresponding $C^*$-algebras. Thus we use the structure of a groupoid as a commutative space on the one hand, which allows us to construct embeddings, and constructions similar to Morita theory on the other hand to construct an equivalent notion of embedding, that is readily stated in terms similar to Morita theory, such that it can be applied to $C^*$-algebras. The resulting procedure will be the noncommutative version of constructing an embedding by embedding a vector bundle into a larger vector bundle and recovering the embedding using the projection in the full vector bundle.

\section{Induced Representations Explained using Quantum Mechanics}

In this section, we will illustrate the conceptual foundations of Morita theory and Rieffel induction \cite{rieffel1974,rieffel1982} on the example of ordinary quantum mechanics in one dimension. We will use the standard notation of physics and collect the corresponding mathematical objects occurring in the example only in the second subsection. The intuition that we gained in this procedure will then be used to propose two strategies for tackling the problem of quantum reduction.

\subsection{Induced Representations for Quantum Mechanics}

Form a physicists viewpoint, the central observation that underlies the theory of strong Morita equivalence is close, but not equivalent, to the fact that one can reconstruct a quantum-algebra from ket-bra operators $|\psi \rangle \langle \phi |$, which are constructed from two ''tame'' Hilbert space vectors $\psi,\phi$. Let us consider how this works in one-dimensional quantum mechanics:
\\
For this purpose let us first revisit the Weyl-algebra of one-dimensional quantum mechanics: It is based on the fundamental Poisson bracket $\{p,x\}=1$. Let us consider the configuration variables, these are elements of $C^\infty_c(\mathbb R)$, i.e. smooth functions of compact support on $\mathbb R$. We can consider the exponentiated Poisson action of the momenta on the configuration variables: $e^{ b \{p,.\}}$, which act as translations of length $b$ on the configuration variables. Moreover, one can consider the special configuration variables $e^{i a x}$, from which we can construct any configuration variable. 
\\
Using these preparations, Weyl quantization is based upon two unitary representations of the groups of Weyl-operators $U(a)=e^{i a x}$ and $V(b)=e^{i b p}$ of exponentiated position and momentum operators, which satisfy the exponentiated Heisenberg commutation relations:
\begin{equation}
  U^*(a) V^*(b) U(a) V(b) = e^{i ab}.
\end{equation}
These two families have a representation on the continuous functions of compact support on $\mathbb R$ by mapping a function $f \in C_c(\mathbb R)$:
\begin{equation}
  \begin{array}{rcl}
    U(a) f : x & \mapsto & e^{iax} f(x)\\
    V(b) f : x & \mapsto & f(x-b).
  \end{array}
\end{equation}
This action can be extended to the Hilbert-space $L^2(\mathbb R,dx)$ of functions on $\mathbb R$ which are square integrable with respect to the Lebesgue measure on $\mathbb R$, by density of $C_c(\mathbb R)$ in $L^2(\mathbb R,dx)$. The Weyl-algebra of quantum mechanics can then be constructed through the Weyl-quantization procedure: Given a continuous function $F$ of compact support on the phase space $\Gamma=\mathbb R^2$, then one can associate an operator to it using the Weyl-correspondence:
\begin{equation}
  \hat{F} := \int da db \tilde{F}(a,b) U(a) V(b),
\end{equation}
where $\tilde{F}$ denotes the Fourier transform of $F$. This action is naturally defined on $C_c(\mathbb R)$ through the action of the families of unitary operators $U$ and $V$ and it can again be extended to a Hilbert space representation by density of $C_c(\mathbb R)$ in $L^2(\mathbb R,dx)$, i.e. the canonical Schr\"odinger representation of QM. The completion of these Weyl quantized operators $\hat F$ in the Hilbert space norm yields a $C^*$-algebra, where the involution is given by the adjoined operation in the Hilbert-space. 
\\
Let us first consider the commutative algebra of configuration observables, which are functions on the phase space, which are independent of the momenta: For any function $f$ of compact support on the configuration space $\mathbb R$, we associate the operator:
\begin{equation}
  \alpha(f) := \int da dx \frac{e^{-iax}}{2 \pi} f(x) U(a).
\end{equation}
On the other hand, we have the group of exponentiated momenta $V(b)=e^{ibp}$. We can now state the Heisenberg commutation relations for configuration variables $f(x)$ and exponentiated momenta $V(b)$:
\begin{equation}
  V^*(b) \alpha(f) V(b) = \alpha(x \mapsto f(x-b)).
\end{equation}
It turns out to be convenient to consider functions $G\in \mathbb R \times \mathbb R$ of compact support on the product of configuration space (first copy of $\mathbb R$) with the group of exponentiated momenta (second copy of $\mathbb R$) and associate the Weyl-operator: 
\begin{equation}
  \pi (G) := \int db \alpha\biggl(x \mapsto G(x,b)\biggr) V(b),
\end{equation}
where we use the fact, that $G$ is a function depending only on the configuration space for each fixed value of $b$. The norm-completion of the Weyl-quantization of these functions is the Weyl-algebra of ordinary quantum mechanics.
\\
Let us now consider, how we can use the module $C_c(\mathbb R)$ to calculate this Weyl-algebra using the ket-bra operators. Evidently, given two elements $f_1,f_2 \in C_c(\mathbb R)$, we can construct an operator $O_{f_1,f_2}$ without using a particular Hilbert-space representation of the families $U,V$ by:
\begin{equation}
  O_{f_1,f_2}:= \pi \biggl((x,b) \mapsto f_1(x) \overline{f_2(x-b)}\biggr).
\end{equation}
In order to show, that the operators $O_{f_1,f_2}$ span a dense subset of the Weyl-algebra, we consider an approximate identity $Id_{C,U(e),\epsilon}$\cite{reiffel1982II}, which is indexed by compact subsets $C$ of $\mathbb R$, neighbourhoods $U(e)$ of the unit element of the group of momentum Weyl-operators $V$ and $\epsilon \in \mathbb R^+$ with the following properties:
\begin{equation}
  \begin{array}{rcl}
    Id_{C,U(e),\epsilon}(x,b) &= 0 & {\textrm{ for }b\textrm{ outside }} U(e)\\
    |Id_{C,U(e),\epsilon}(x,b)-1| &< \epsilon & {\textrm{ for }} x \in C.
  \end{array} 
\end{equation}
To each such a function $Id$, we associate its Weyl-quantization $\pi(Id)$, which defines an approximate identity in the Weyl-algebra. Let us now provide functions $Id$, which are sums of the operators $O_{f_1,f_2}$, that satisfy these properties: For each compact subset $C$ of $\mathbb R$, we can find a covering of $C$ by open sets $U_i$, such that the size of these sets is smaller than some $\delta >0$. Now for every $\epsilon > 0$, we can find a regularization of the characteristic function of $C$, which differs at most by $\epsilon$ form 1 in $C$, which is the sum of continuous positive functions $\chi^\epsilon_i$ with respective support on $U_i$. For each neighbourhood $U(e)$ of the identity in the group of exponentiated momenta, we can choose a positive $\delta$, that is smaller than half the size of $U(e)$. Then the operator $O_{\chi^\epsilon_i,\chi^\epsilon_i}$ satisfies the second condition of the approximate identity. Thus their sum
\begin{equation}
  Id_{C,U(e),\epsilon}=\sum_i O_{\chi^\epsilon_i,\chi^\epsilon_i}
\end{equation}
yields the desired approximate identity, which can now be used to see, that for any $G \in C_c(\mathbb R \times \mathbb R)$ there is a sequence of operators $O_{f_1,f_2}$, such that the associated Weyl-operators converge to $\pi(G)$, since $\pi(G)Id \rightarrow \pi(G)$ and on the other hand
\begin{equation}
  \pi(G) \pi(Id)= \pi(G) \sum_i O_{\chi^\epsilon_i,\chi^\epsilon_i} = O_{\pi(G)\chi^\epsilon_i,\chi^\epsilon_i} \rightarrow \pi(G),
\end{equation}
where $\pi(G)$ acts on $\chi^\epsilon_i$ by the Schr\"odinger action $C_c(\mathbb R)$. The extension to the entire Weyl-algebra is defined by density of the operators $\pi(G)$. Thus, we have reconstructed the Weyl-algebra of QM as the norm closure of the ''rank one operators'' $O_{f_1,f_2}$.
\\
Let us summarize, what we have done: We used a non-Hilbert representation of quantum mechanics on the continuous functions of compact support on the configuration space $C_c(\mathbb R)$ and we defined a bilinear structure $O$ with values in the Weyl-algebra on it by setting: $O:(f_1,f_2)\mapsto O_{f_1,f_2}$. The span of this bilinear structure then turned out to dense in the Weyl-algebra, which we saw using the approximate identity trick. The mathematical structure, that underlies this construction are pre-$C^*$-modules:  $C_c(\mathbb R)$ together with $O$ turns out to be such a pre-$C^*$-module. On the mathematical side, there is a close similarity between the bilinear structure, which is $O$ in our example, and Hermitian inner products: If we had started out with the mathematical definition of a pre-$C^*$-module, then we would have seen, that an inner product is just the special case of such a bilinear structure, that arises when the range of the bilinear structure is the rather small algebra $\mathbb C$ of complex numbers and not the Weyl-algebra as in our example.
\\
Morita theory now comes into play through a very simple observation: The operators $O_{f_1,f_2}$ act on a function $f \in C_c(\mathbb R)$ through the canonical representation. Thus, we can define a second set of ''rank one'' operators $T_{f_2,f}$ through the formula $ f_1 T_{f_2,f} := O_{f_1,f_2} f $, which is again represented as a set of linear transformations on $C_c(\mathbb R)$, by:
\begin{equation}
  T_{f_2,f}: f_1 \mapsto O_{f_1,f_2} f.
\end{equation}
Using the formula for $O_{f_1,f_2}$ from above, we obtain:
\begin{equation}
  T_{f_2,f} f_1: x \mapsto f_1(x) \int db \overline{f_2(x-b)} f(x-b)= f_1(x) \int db \overline{f_2(b)} f(b),
\end{equation}
from which we can read off, that the operators $T_{f,g}$ take values in the complex numbers $c=\int db \overline{f_2}(b) f(b)$. In mathematical language: $C_c(\mathbb R)$ is also a $C^*$-module for $\mathbb C$. The importance of this result lies in the availability of a construction, called Rieffel induction, that allows one to use this $C^*$-bi-module structure of $C_c(\mathbb R)$ to construct a Hilbert space representation of the Weyl-algebra out of a given Hilbert-space representation of the complex numbers using a slight change of notation from the GNS construction of a representation out of a given ''vacuum state'':
\\
Let us consider the following, very simple and in fact unique vacuum state on $\mathbb C$:
\begin{equation}
  \omega(c)=c,
\end{equation}
which obviously satisfies the state conditions $\omega(c^*c) \ge 0$, $\omega(1)=1$ and $\omega(\lambda c) = \lambda c$. Now, we can use the rank one operators $T_{f,g}$ with values in $\mathbb C$ to define a vacuum state $\Omega$ on the Weyl-algebra by the following formula:
\begin{equation}
  \Omega(O_{f,g}) := \omega(T_{f,g}),
\end{equation}
out of which we can preform a GNS-construction to obtain a representation of the Weyl-algebra. The induced vacuum state $\Omega$ is in our case given by:
\begin{equation}
  \Omega(O_{f,g}) = \omega\biggl(\int db \overline{f(b)} g(b)\biggr) =\int db \overline{f(b)} g(b).
\end{equation}
In order to construct the GNS representation, we have to factor out the Gelfand ideal, i.e. the set of positive operators, for which $\Omega$ vanishes. The positive operators are all of the form $O_{f,f}$, such that the only element of $C_c(\mathbb R)$ for which $\Omega(O_{f,f})=\int db |f(b)|^2$ vanishes is $f: x \mapsto 0$, which is the sole element in the Gelfand ideal $\mathfrak J=\{f: x \mapsto 0\}$. This means, that the map $\Pi$ that maps $C_c(\mathbb R)$ into $C_c(\mathbb R)/ \mathfrak I=C_c(\mathbb R)$ is trivial: $\Pi: f \mapsto f$.
\\
The GNS construction now tells us, that the completion of $C_c(\mathbb R)$ in the induced inner product defined by:
\begin{equation}
  \langle f,g \rangle_{ind} := \Omega(O_{\Pi(f),\Pi(g)})= \int db \overline{f(b)}g(b),
\end{equation}
is a Hilbert-space, which carries the induced representation $\pi_{ind}$ of the Weyl-algebra by:
\begin{equation}
  \pi_{ind}(A) \Pi(f) := \Pi(\pi(A)f)=\pi(A)f,
\end{equation}
where $\pi$ denotes the canonical representation of the Weyl-algebra on $C_c(\mathbb R)$. Thus, the Hilbert-space completion of $C_c(\mathbb R)$ in $\langle.,.\rangle_{ind}$ turns out to be $L^2(\mathbb R,dx)$ the fundamental Schr\"odinger representation of the Weyl-algebra. 
\\
One of the benefits of Morita theory is, that it defines an equivalence relation: Given a $C^*$-bi-module, that satisfies certain mathematical properties, one can define an inverse ''conjugate $C^*$-bi-module'' by basically reversing the role of the operators $O$ and $T$, which act on the complex conjugate by the complex conjugate action. It turns out that if a state $\omega^\prime$ on one of the two involved algebras that was induced out of a state $\Omega$ that has been induced from say $\omega$ using the conjugate $C^*$-bi module, then $\omega^\prime$ is equivalent to $\omega$. Thus, following Morita theory, we have just outlined a proof of the Stone-von Neumann theorem: Given any state of the Weyl-algebra of quantum mechanics, we can induce a state on the complex numbers, which just have one irreducible representation, corresponding to the state $\omega: c \mapsto c$, which by the induction above is the Schr\"odinger representation.
\\
Before we take a step back and look at the involved mathematical structure, we want to show, that both the Weyl-algebra and $\mathbb C$ are full subalgebras of a larger algebra, that can be constructed using the bi-module $C_c(\mathbb R)$: Take any element $\pi(G)$ of the Weyl-algebra, a complex number $c$, a function $f \in C_c(\mathbb R)$ and a function $\bar g \in \bar{C}_c(\mathbb R)$, where $\bar C_c(\mathbb R)$ denotes the conjugate bi-module, then we can form a matrix:
\begin{equation}
  \biggl(
    \begin{array}{cc}
      \pi(G) & f \\
      \bar g & c
    \end{array}
  \biggr)
\end{equation}
These matrices have a natural product, which can be quoted by using the formulas for the two families of rank one operators $O,T$:
\begin{equation}
  \biggl(
    \begin{array}{cc}
      \pi(G_1) & f_1 \\
      \bar g_1 & c_1
    \end{array}
  \biggr)
  \biggl(
    \begin{array}{cc}
      \pi(G_2) & f_2 \\
      \bar g_2 & c_2
    \end{array}
  \biggr)
=  \biggl(
    \begin{array}{cc}
      \pi(G_1)\pi(G_2)+ O_{f_1,g_2} & \pi(G_1)f_2 + c_2 f_1 \\
      \bar g_1 \pi(G_2) + c_1 \bar{g}_2 & T_{g_1,f_2} + c_1 c_2
    \end{array}
  \biggr).
\end{equation}

\subsection{Mathematical Structures}

Let us now extract the underlying mathematical structure: The Weyl-algebra of QM was constructed to be a $C^*$-algebra. The representation space $C_c(\mathbb R)$ is a complex module with a continuous action of the Weyl-algebra by the fundamental representation. The ''rank-one'' operators $O_{f,g}$ on the other hand defined a bi-linear structure on this module, which is dense in the Weyl-algebra, and which satisfies properties, that are analogue to the ones of an inner product: $O_{f,g}^*=O_{f,g}$, $O_{f,f} \ge 0$ and $O_{f,f}=0$ only if $f=0$ as well as the compatibility with the action of the Weyl-algebra $O_{f,g\pi(A)}=O_{f,g}\pi(A)$. This is the mathematical definition of a {\bf Hilbert $C^*$-module}. 
\\
Form this Hilbert-$C^*$-module we where able to construct an ''{\bf induced $C^*$-algebra}'' $C^*(Weyl,C_c(\mathbb X))$ as the completion of the rank one operators obtained as a compact completion of the linear transformations $h T_{f,g}:= O_{h,f} g$ on the module $C_c(\mathbb R)$. It turns out, that the same construction is always possible. Using a slight generalization of the GNS-construction, we obtained {\bf Rieffel induction}, which turns out to be a functor, which takes states and hence representations over one $C^*$-algebra to states over the induced $C^*$-algebra, which actually turns out to be an equivalences of categories between the unitary equivalence classes of continuous representations of the original and the induced $C^*$-algebra. This is the reason, why the module is called a Morita-equivalence bi-module between the two Morita equivalent $C^*$-algebras.
\\
Now, given two Morita equivalent $C^*$-algebras $\mathfrak A,\mathfrak B$ and an given the mediating equivalence bi-module, we can always construct a linking algebra of matrices:
\begin{equation}
    \biggl(
    \begin{array}{cc}
      A & f \\
      \bar g & B
    \end{array}
  \biggr)
\end{equation} 
and declare their matrix product using the bi-linear structure on the equivalence bi-module and its conjugate, where $A$ and $B$ are elements of the respective $C^*$-algebras $\mathfrak A$ and $\mathfrak B$ and $f,\bar g$ are elements of the equivalence bi-module and its conjugate. In this way, we can always construct a matrix algebra, such that the two $C^*$-algebras are full\footnote{A subalgebra $\mathfrak A$ of $\mathfrak B$ is called full, if the inverse of an element of $\mathfrak A$ exists in $\mathfrak B$, then it is in $\mathfrak A$.} and hereditary\footnote{A subalgebra $\mathfrak A$ of $\mathfrak B$ is called hereditary, iff elements of the form $\mathfrak A \mathfrak B \mathfrak A$ are in $\mathfrak A$.} subalgebras. The converse is also true: If there exists a $C^*$-algebra $\mathfrak C$, such that both $\mathfrak A$ and $\mathfrak B$ are full hereditary subalgebras, then $\mathfrak A$ and $\mathfrak B$ are Morita equivalent. The Morita equivalence bimodule is $\mathfrak {ACB}$, i.e. the ideal of elements $\{acb: a \in \mathfrak A, c \in \mathfrak C, b \in \mathfrak B\}$.
\\
Now, as we saw the mathematical structure involved, we can ask ourselves, whether there are generalizations to our quantum mechanics example, such that we can repeat the above construction. The data, that we constructed the Weyl-algebra from was a configuration space, given by a manifold $\mathbb X$. The configuration variables that we considered where the algebra $C_c^\infty(\mathbb X)$ of smooth functions of compact support on the configuration space. Then we used the exponentiated Poisson action of a set of momenta $p$: $e^{i  \{p,.\}}$ on the algebra of configuration variables and obtained a smooth action of the a Lie-group of momentum transformations on the configuration manifold, which extended to a ''unitary'' action of the momentum group on the algebra of configuration variables as pull-backs under these translations. In turn, given a smooth action $\alpha$ of a Lie-group $G$ on a manifold $\mathbb X$, we can abstractly consider $\mathbb X$ as the configuration space and $G$ as the momentum group, corresponding to the Poisson action of the momenta, that is given by the Lie-action of the Lie-algebra $\mathfrak g$ of $G$ on the configuration space $\mathbb X$. This is indeed the type of structure, that we will base our notion of quantization upon.
\\
Moreover, if the group action of $G$ on $\mathbb X$ is free, then we can again define a faithful unitary action of $G$ on the algebra of configuration variables $C_c(\mathbb X)$ and apply the analogue to Weyl-quantization as described above. In addition, it will turn out, that we can turn $C_c(\mathbb X)$ into a bi-module over the quantum algebra $C^*(\mathbb X,G)$ and over the commutative algebra $C(\mathbb X/G)$ of functions on the $G$-orbits in $\mathbb X$. The two sets of bilinear operators are given by the completely analogous formulas: (1) $O_{f_1,f_2}:= \pi((x,g) \mapsto f_1(x)\overline{f_2(\alpha_g(x))}g)$ and (2) $T_{f_1,f_2}:= ([x]_G \mapsto \int_G d\mu_h(g) \overline{f_1(\alpha_g(x))} f_2(\alpha_g(x)))$.

\subsection{Two Strategies}

The goal of this paper is to provide a construction, that is the quantum analogue to the ''pull-back'' under a Poisson-embedding. With the mathematical preparations in the previous sections, there are two obvious ways to proceed. The first is related to the idea of group averaging, which in turn turns out to be closely related to the linking algebra. The second is more directly related to $C^*$-modules: It relies on the observation, that a Hermitian vector bundle\footnote{A Hermitian vector bundle is a complex vector bundle with a Hermitian inner product on each fibre\cite{rieffel1982II}.} over a space $\mathbb X$ is nothing else than a $C^*$-module over the commutative $C^*$-algebra $C(\mathbb X)$. Since any embedding of a space $\mathbb Y$ into a space $\mathbb X$ induces a pull-back under this embedding from the trivial Hermitian vector bundle $C(\mathbb X)$ with fibre-wise inner product $\langle f,g \rangle_x := \overline{f(x)}g(x)$ to the trivial Hermitian vector bundle $C(\mathbb Y)$, we will express this pull-back in terms of this trivial vector bundle and generalize it to the noncommutative case, where the vector bundle is replaced by a $C^*$-module, which acts as its noncommutative counterpart.

\subsubsection{Linking Algebra Approach}

Let us start with the embedding $\eta$ of a space $\mathbb X_o$ into another space $\mathbb X$. This embedding acts as by its pullback on the commutative $C^*$-algebras $C(\mathbb X)$:
\begin{equation}
  \eta: C(\mathbb X) \rightarrow C(\mathbb X_o): (x \mapsto f(x)) \mapsto (x_o \mapsto f(\eta(x_o))).
\end{equation}
In turn, we can view $\eta^*$ as a map $P$, that maps $C(\mathbb X)$ into the equivalence classes of functions, that differ only by an element of the ideal $\mathfrak I$ of functions, that vanish on $\mathbb X_o$:
\begin{equation}
  P: C(\mathbb X) \mapsto C(\mathbb X) / \mathfrak I.
\end{equation}
If we use a physical point of view, that $\mathbb X_o$ is the space, that is accessible to our measurements, such that $C(\mathbb X)$ represents the observables, that our measurements are sensitive to, we can consider an element of the ideal $\mathfrak I$ as a gauge transformation, that does not change measurable physical quantities, which are represented by the gauge equivalence classes. Thus, we can enlarge the commutative algebra of functions $C(\mathbb X)$ by adding a gauge group $H$ of transformations, which is given by the action of the commutative group $(\mathfrak I,+)$. The idea is basically to construct a noncommutative algebra, that is Morita equivalent to $C(\mathbb X_o)/\mathfrak I$. 
\\
Such a $C^*$-algebra can be constructed, if we find a group $G$ together with a free and proper action on $\mathbb X$, such that $\mathbb X /G$ is isomorphic to $\mathbb X_o$, then we have a noncommutative $C^*$-algebra, called the transformation group algebra $C^*(\mathbb X,G)$, that encodes this $C(\mathbb X_o)$ in the sense, that $C(\mathbb X_o)$ is the unique commutative $C^*$-algebra, that is Morita equivalent to the transformation group $C^*$-algebra.\footnote{We will discuss transformation group $C^*$-algebras later in this paper.} Given the group $G$, we see immediately, that it is the group, that we would like to use to preform group-averaging in order to obtain $C(\mathbb X_o)$ form $C(\mathbb X)$.
\\
Obviously, unless the set $\eta(\mathbb X_o)$ is open in $\mathbb X$, which is a rather restrictive case, there are no continuous functions $f \in C(\mathbb X)$, that span a subalgebra of $C(\mathbb X)$, that is isomorphic to $C(\mathbb X_o)$, rather $C(\mathbb X_o)$ is a quotient by the ideal $\mathfrak I$ of functions, that vanish on the closure of $\eta(\mathbb X_o)$. Thus, although there is a natural projection $P$, there is in general no algebra-morphism, that embeds $C(\mathbb X_o)$ into $C(\mathbb X)$. However, there are many linear subspaces, such that embeds $C(\mathbb X_o)$ embeds as a linear subspace into $C(\mathbb X)$. 
\\
Taking these linear subspaces as representatives, we can in a certain sense view the transformation group algebra $C^*(\mathbb X,H)$ as a linking algebra, that links all these linear subspaces. The relation between $C^*(\mathbb X,G)$ and $C^*(\mathbb X,H)$ will become clearer, once we consider the associated transformation groupoids, but this is beyond the scope of this paper. The focus of this paper will be on another approach to the problem of finding a reduced quantum algebra, which is inspired by Hermitian vector bundles.

\subsubsection{Hermitian Vector Bundle Approach}

The idea behind this approach is to generalize the construction of an embedded space by embedding a fibre bundle over this space into a larger vector bundle over a larger space using a bundle morphism and then using the projection in the larger vector bundle to obtain the embedded space by the restriction to the embedded fibre bundle. Although this procedure is a very odd and complicated way of constructing an embedding for a space, it turns out that this procedure can be applied also for noncommutative spaces, because there certain noncommutative vector bundles, which can be thought of as functions over commutative spaces, such that we are able to use the ordinary embedding of these underlying spaces to apply our construction.
\\
First, the Serre-Swan theorem tells us that we can think of finitely generated projective modules over $C(\mathbb X)$ as sections in a vector bundle over $\mathbb X$. On a complex vector bundle, we can introduce a Hermitian structure $(.,.)_x$ in each fibre which acts as fibering the module over $\mathbb X$, which results in considering Hermitian modules over $C(\mathbb X)$. The $C^*$-modules that we encountered in Morita theory are Hermitian modules, when the underlying algebra is commutative, which allows us to consider them as noncommutative vector bundles.
\\
Let us be more specific: We can use the canonical trivial Hermitian vector bundle $C_c(X)$ over a space $\mathbb X$, which has the canonical Hermitian structure $\langle f,g \rangle_x = \overline{f(x)} g(x)$. Let us now consider an embedding $\eta: \mathbb X_o \rightarrow \mathbb X$: It defines a Hermitian vector bundle over $\mathbb X_o$ through, which is $\eta^*C(\mathbb X)=C(\mathbb X_o)$ with the induced Hermitian structure:
\begin{equation}
  \langle \eta^*f,\eta^*g \rangle_{x_o} := \overline{f(\eta(x_o))}g(\eta(x_o)).
\end{equation}
Now using the ''projection'' $P: f \mapsto \eta^* f$, we can define a Hermitian structure on $C(\mathbb X)$ with values in $C(\mathbb X_o)$ by:
\begin{equation}
  \langle f,g \rangle_{x_o}= P(\langle f,g \rangle_\mathbb X)_{x_o}.,
\end{equation}
where $\langle .,. \rangle_{\mathbb X}=(x \mapsto \langle .,. \rangle_x)$ denotes the Hermitian structure on $C(\mathbb X)$. We observe, that only using $P$, we can transform the Hermitian vector bundle $C(\mathbb X)$ into a Hermitian vector bundle over $\mathbb X_o$. Using the induced Hermitian structure $P(\langle.,.\rangle_{\mathbb X})$, we can immediately calculate the embedded space $\mathbb X_o$ from the full vector bundle $C(\mathbb X)$ as the Gelfand spectrum of the image of $x_o \mapsto \langle .,. \rangle_{x_o}$. 
\\
Hermitian vector bundles on the other hand are $C^*$-modules for the special case, that the underlying algebra is a commutative $C^*$-algebra and hence isomorphic to $C(\mathbb X)$ for some locally compact $\mathbb X$. This suggests an obvious strategy: 
\\
Take a $C^*$-module over a $C^*$-algebra and define a map $P$ analogue to the projection $P$ in the commutative case, such that we are able to calculate the ''embedded'' $C^*$-algebra as the image of $P$ using the same method, that we used to calculate the Weyl-algebra of QM in the previous example. It is however obvious, that the occurring modules can in general not be equivalence bi-modules for a very simple reason: Consider the commutative case, and embed a space $\mathbb X$ into the product with another space $\mathbb Y$, then if $\mathbb Y$ is a nontrivial topological space, $C(\mathbb X)$ will not be Morita equivalent with $C^*(\mathbb X\times \mathbb Y)$. Thus, we will call the occurring bi-modules embedding bi-modules and we need to consider slight modifications of the methods used in Morita theory. 
\\
In the previous example we showed, that one can use Rieffel induction to induce a Hilbert space representation of a Morita equivalent algebra out using the bi-module. We will later use a construction very similar to Rieffel induction, to induce a Hilbert space representation of the embedded $C^*$-algebra using the embedding bi-module.

\section{Groupoids as mediating Structures in Quantization}

The notion of quantization, that we want to use in this section is heavily based on transformation group systems, that arise naturally in physics, whenever a group of exponentiated momenta acts freely as a group of transformations on smooth functions on a configuration space through a pull-back under a group of homeomorphisms on the configuration space. This structure is most easily encoded in a Lie-transformation groupoid.\footnote{This section follows ideas from \cite{muhly}.}
\\
In this section we will see that Lie-groupoids in general and transformation-Lie-groupoids in particular play the role of a mediating structure in the sense, that there is a unique classical Poisson system associated with each Lie-groupoid and on the other hand that there is a unique associated quantum system (i.e. a $C^*$-algebra).

\subsection{Transformation Groupoids}

Before we change our viewpoint of the Weyl-group of exponentiated momenta, that acts on a configuration space to a transformation groupoid, we will first review general notions about groupoids\cite{connes}:
\\
A groupoid is a small category in which each morphism is invertible. Thus, a groupoid contains of a set $\mathcal G_o$, called the unit set, which contains the objects of the small category. Secondly it contains of a set $\mathcal G$, whose elements are called groupoid elements, which is the set of morphisms in the small category. By assigning the identity morphism of each group element, we obtain a map $e:\mathcal G_o \rightarrow \mathcal G$, called the unit map. Moreover, using the domain and codomain for each morphism, we obtain two maps $r,s:\mathcal G \rightarrow \mathcal G_o$, called the source and range map. The set $\mathcal G^2=\mathcal G * \mathcal G:=\{(g_1,g_2)\in \mathcal G\times \mathcal G: r(g_1)=s(g_2) \}$ consists of the composable pairs of morphisms and $g^{-1}$ denotes the inverse of a given morphism $g\in \mathcal G$. Using these notations, the categorial axioms read:
\begin{enumerate}
  \item $s(g_1\circ g_2)=s(g_2)$ for all $(g_1,g_2)\in \mathcal G^2$
  \item $r(g_1\circ g_2)=r(g_1)$ for all $(g_1,g_2)\in \mathcal G^2$
  \item $s(e(x))=x=r(e(x))$ for all $x \in \mathcal G_o$
  \item $g \circ e(s(g))=g=e(r(g))\circ g$ for all $g \in \mathcal G$
  \item $(g_1 \circ g_2)\circ g_3=g_1 \circ (g_2 \circ g_3)$ for all $g_1,g_2,g_3 \in \mathcal G$
  \item $g^{-1}g=e(r(g))$ and $g g^{-1}=e(s(g))$ for all $g \in \mathcal G$
\end{enumerate}
A particular, however not general, example of a groupoid is the semi-direct product of a group $G$ acting by an action $\alpha$ on a set $\mathbb X$:
\\
The groupoid set is $\mathcal G=\mathbb X \times G$, the unit set is $\mathcal G_o=\mathbb X \times \{e\}$ and the set of composable elements $\mathcal G^2=\{((x_1,g_1),(x_2,g_2))\in \mathcal G \times \mathcal G:x_2=\alpha_{g_1}(x_1)\}$. The groupoid structure on this data is:
\begin{enumerate}
  \item $e: x \mapsto (x,e)$ for all $x\in \mathcal G_o$
  \item $r(x,g)=x$ for all $(x,g) \in \mathcal G$
  \item $s(x,g)=\alpha_g(x)$ for all $(x,g) \in \mathcal G$
  \item $(x_1,g_1)\circ(x_2,g_2)=(x_1,g_1g_2)$ for all $((x_1,g_1),(x_2,g_2)) \in \mathcal G^2$
  \item $(x,g)^{-1}=(\alpha_g(x),g^{-1})$
\end{enumerate}
If the group action $\alpha$ is free and proper, then we call the groupoid structure associated to this semi-direct product a {\bf free transformation groupoid}.
\\
By considering a group as a groupoid whose unit space consists of $\{e\}$, we see, that a groupoid is the generalization of a group that allows many units. This is the reason why we will have to introduce a momentum map, when we want to generalize a group action on a space to the action of a groupoid on a space, which is precisely the map, that tells us, which unit element in the groupoid we have to start with for a given element of the space upon which the groupoid acts.
\\
For the purpose of our construction, we will assume that the transformation groupoids that we consider are smooth, i.e. both $\mathcal G$ and $\mathcal G_o$ are smooth manifolds and $r,s,e,\circ$ are smooth maps. Thus, we assume, that the transformation group data consists of a smooth manifold $\mathbb X$ together with a smooth action $\alpha$ of a Lie-group $G$. We will refer to these groupoids as {\bf Lie-transformation-groupoids} and we will particularly be interested in free Lie-transformation-groupoids. The reason for restricting ourselves to such smooth systems is because we will consider the group as a Weyl-group of the exponentiated Poisson action of classical momenta, which acts smoothly on a space of smooth functions on a configuration space. This ensures that we are able to take the classical limit.

\subsection{Associated Classical System}

We will now construct a classical Poisson system. For this purpose, we will first introduce Lie-Algebroids, which are classically a dual description of a Poisson system. The name suggests rightfully, that a Lie-algebroid describes the tangent structure of a Lie-groupoid at the unit space, which is true, however not the general definition of a Lie-algebroid, which is defined as follows\footnote{For details about the relation of Lie-algebroids and Poisson systems see e.g. \cite{cannas-weinstein,mackenzie-groupoid,vaismann}.}:
\begin{defi}
  A {\bf Lie-algebroid} is a vector bundle $(E,\pi,\mathbb X)$ over a manifold $\mathbb X$, together with a Lie-bracket $[.,.]_E$ defined on smooth sections $\Gamma^\infty(E)$ and a morphism $\rho$ of vector bundles from $E$ to the tangent bundle $T(\mathbb X)$, called the anchor map, which for all $s_1,s_2\in \Gamma^\infty(E)$ and for all $f \in C^\infty(\mathbb X)$ satisfies a generalization of Leibniz rule:
  $$
    [s_1,f s_2]_E = f [s_1,s_2]_E + (\rho(s_1)f)t,
  $$
where $\rho(s)$ denotes the section $\rho\circ s$ in $T(\mathbb X)$.
\end{defi}
Let us now explain, how we are able to build a Poisson system from a general Lie-algebroid: For this purpose, let us consider the dual $E^*$ of $E$, from which we construct symmetric tensor products of sections in the dual bundle, which are naturally isomorphic to polynomials in the sections in $E^*$ using the canonical isomorphism $I$. Moreover, we can associate a smooth function $f \in C^\infty(\mathbb X)$ as a polynomial $\pi^* f$ of degree zero in the dual vector fields using the projection $\pi$ in $E$, such that we obtain an algebra of polynomials, on which we define a Poisson bracket $\{.,.\}$ by:
\begin{equation}
  \begin{array}{rcl}
    \{\pi^* f_1,\pi^*f_2\} & := & 0\\
    \{\pi^* f,I(s)\} & := & \pi^*(\rho(s)f)\\
    \{I(s_1),I(s_2)\} & := & -I([s_1,s_2]),
  \end{array}
\end{equation}
where $f,f_1,f_2 \in C^\infty(\mathbb X)$ and $s,s_1,s_2 \in \Gamma^\infty(E)$. The Poisson bracket is then extended by Leibniz-rule and bilinearity to the entire polynomial algebra.
\\
The inversion of this procedure also exists, such that we are able to associate a Lie-algebroid structure on $E$ to any Poisson system on $E^*$. What we need to complete the construction of a classical system is to associate a Lie-algebroid to a given Lie groupoid: This is done exactly in the same way as a Lie-algebra is associated to a Lie-group: One considers the algebra of left invariant vector fields by considering them as a Lie-algebra on the tangent space at the unit element, which is determined by the derivative action of the left-invariant vector fields. The only difference is that we ''have many units'', i.e. we have the entire configuration space $\mathbb X$ over which a Lie-algebra is fibered as vector fields. This naturally defines $\rho(E)$ as vector fields on $T(\mathbb X)$. The abstract Lie-algebra is then denoted by $E$ and the anchor map $\rho$ is recovered as the natural map from $E$ to the tangent bundle $T(\mathbb X)$.
\\
Let us consider this more specifically for a transformation groupoid and associate its transformation Lie-algebroid:
\\
Consider a transformation groupoid given by a manifold $\mathbb X$ upon which a Lie-group $G$ acts by a smooth action $\alpha$, which is assumed to be free and proper. Then the derivative action $\phi$ of the Lie-algebra $\mathfrak g$ of $G$ on $T(\mathbb X)$, given by the collection of the derivative actions at each point in $\mathbb X$ defines an action $\phi: \mathfrak g \rightarrow \Gamma^\infty(T^*(\mathbb X))$. Since the groupoid is $\mathbb X \times G$, we obtain a trivial fibre bundle $E=\mathbb X \times \mathfrak g$. We are able to calculate the Lie-derivative $\mathcal L_X s$ of a section $s \in \Gamma^\infty(E)$ along a vector field $X$ by defining the derivative component-wise, since the vector bundle is trivial.
\\
Moreover, we are able to define the anchor map $\rho: E \rightarrow T(\mathbb X)$ by:
\begin{equation}
  \rho(x,l):=\phi(l)|_x,
\end{equation}
where $x \in \mathbb X$ and $l \in \mathfrak g$, such that $(x,l) \in E$. Using this anchor map, we can express the Lie-algebroid structure $[.,.]_E$, which arises as the Lie-bracket of the derivative action of the sections in the trivial principal bundle $\mathbb X \times \mathbb G$ as:
\begin{equation}
  [s_1,s_2]_E(x)=[s_1(x),s_2(x)]_{\mathfrak g} + (\mathcal L_{\phi(s_1)}s_2)(x) - (\mathcal L_{\phi(s_2)}s_1)(x),
\end{equation}
where $s_1,s_2 \in \Gamma^\infty(E)$ and $[.,.]_{\mathfrak g}$ is the Lie-bracket in $\mathfrak g$. The aforementioned extension to all polynomials of smooth sections in $E$ defines a Poisson system. Thus, we can identify this procedure as the inverse of the procedure of taking the exponentiated Poisson action of the momenta and letting this group act as pullbacks under finite group-translations on the configuration space. 

\subsection{Associated Quantum System}

In the previous section, we constructed a Poisson system from a given Lie-transformation groupoid. In this section, we will show, how one can construct a $C^*$-algebra form a transformation groupoid $\mathcal G(\mathbb X,G)$, which we will interpret as the Weyl-algebra of quantum observables associated to this system.
\\
The $C^*$-algebra that we want to construct is a completion of the convolution algebra, which we already used as the quantum algebra for ordinary quantum mechanics. The trick to become independent of a certain Haar system (although easily constructable for a free transformation groupoid) is to consider $1/2$-densities on the groupoid\cite{connes}. Denoting the set of elements $\mathcal G^x:=\{g \in \mathcal G: r(g)=x\}$ and similarly $\mathcal G_y:=\{g \in \mathcal G: s(g)=y\}$ and the dimension of the fibres $\mathcal G^x,\mathcal G_y$ by $k$, then a $1/2$-density is a section in the bundle over $\mathcal G$, whose fibre at $g: r(g)=x, s(g)=y$ is the linear space of maps $a_g: \wedge^k T_g(\mathcal G^x) \otimes \wedge^k T_g(\mathcal G_y) \rightarrow C$, which satisfy:
\begin{equation}
  a_g(\lambda v)= \sqrt{|\lambda|} \rho(v) \,\,\, \forall \lambda \in \mathbb R.
\end{equation}
The merit of this condition is, that we can integrate $a_1\circ a_2$for two $1/2$-densities $a_1,a_2$  without introducing a measure over $\mathcal G^x$, thus we can use the convolution product:
\begin{equation}
  a_1 * a_2 (g) := \int_{g_1\circ g_2 = g} a_1(g_1)a_2(g_2).
\end{equation}
Using this convolution product for $1/2$-densities, we can define an algebra of smooth compactly supported $1/2$-densities $C_c^\infty(\mathcal G,1/2)$, which has a natural involution given by:
\begin{equation}
  a^* : g \mapsto \overline{a(g^{-1})} \,\,\,\forall g \in \mathcal G.
\end{equation}
For this involution, we can define an involutive Hilbert-space representation of this algebra for each point $x\in \mathcal G_o$ given by:
\begin{equation}
  \pi_x(a) f : g \mapsto \int a(h) f(h^{-1}g) \,\,\,\forall g \in \mathcal G_x,
\end{equation}
where $f$ is an element of the Hilbert-space $L^2(\mathcal G_x)$ of square integrable half-densities on $\mathcal G_x$. Using this family of representations, we can define a norm using the Hilbert norm $\mathcal G_x$:
\begin{equation}
  ||a||:=\sup_{x \in \mathcal G_o} ||\pi_x(a)||.
\end{equation}
It turns out, that the completion of $C^\infty_c(\mathcal G,1/2)$ with the aforementioned involution is a $C^*$-algebra for this norm.
\\
Of course, when restricting ourselves to free transformation groupoids $\mathcal G(\mathbb X,G)$, we have a much simpler construction of a quantum algebra, which occurs rather naturally in physics:
\\
First choose a nondegenerate measure on $\mu_{X/G}$ on $\mathbb X/ G$ and consider the product measure $\mu$ of $\mu_{X/G}$ and the Haar measure $\mu_H$ on $G$: Construct the Hilbert space $\mathcal H:=L^2(\mathbb X,\mu)$ and present a smooth function of compact support on the groupoid $a: (x,g)\mapsto a(x,g)$ on $\mathcal H$ by:
\begin{equation}
  a v : x \mapsto \int d\mu_h(g) a(x,g) v(\alpha_g(x)).
\end{equation}
Second, using the adjoining in the Hilbert-space as an involution yields the same formula for the involution as above and using the completion in the Hilbert-norm yields a $C^*$-algebra, as we know it from quantum mechanics, and the representation that we used in its construction is a Schr\"odinger representation, which we will call the fundamental representation for a given transformation groupoid system. 
\\
Thus, we canonically associated a well known quantum algebra to a given Lie-transformation group system.

\subsection{Classical Limit}

Having a canonical association of a Poisson system on the one hand and a quantum algebra with a Schr\"odinger type representation on the other hand for a given transformation groupoid system, we can talk about taking the classical limit. 
\\
In this paper we use the term classical limit in the sense that given a transformation groupoid $C^*$-algebra, that describes a quantum system, we associate the according transformation Lie-algebroid and its dual, i.e. a Poisson system to it. In short this is explained in the following diagram, which starts at a free Lie-transformation groupoid based on the action $\alpha$ of a Lie-group $G$ on a manifold $\mathbb X$ in the center of the right column:
  \begin{equation}
    \begin{array}{rcccl}
       &  & measure \, \mu_o  & f_1\star f_2(x,g) &  \rightarrow convolution\,prod. \\
       &rep.\,\,on  & \, on \, \mathbb X / G &  & \\
       & L^2_{\mu_o\mu_H}(\mathbb X)  & \longleftarrow ---- & C^*(\mathbb X,G) &  .^*, ||.|| \\
      \hbar& | &  & \uparrow & funct.\,+\,convolution\,prod.  \\
      \downarrow & | &  & \mathcal G(\mathbb X,G) & (x,g)(y,h)|_{y=\alpha_gx}=(x,gh)  \\
      0 & \downarrow &  & \downarrow & \frac{df(x,g)}{dg}\,at\,\,each\,x \in\mathbb X \\
       & \mathcal P(\mathbb X,G) & \longleftarrow ---- & \mathcal A(\mathbb X,G) & f^i p_i(x), [.,.]_{Lie}  \\
       & f(x,p) & polynomials &  & f(x)\\
       & smooth & in\, momenta & &
    \end{array}\nonumber
  \end{equation}
The only non canonical part in this diagram is the choice of measure $\mu_o$ on $\mathbb X/G$, which is the reason why we consider the association of a Poisson system $\mathcal P(\mathbb X,G)$ to a $C^*$-algebra $C^*(\mathbb X,G)$ as a classical limit, since this procedure is invertible. The association of a Poisson system to a particular Hilbert-space representation on is on the other hand not invertible, since even the unitary equivalence classes of regular, irreducible representations are not unique for a general transformation groupoid systems. Only for those, where the locally compact configuration space $\mathbb X$ is isomorphic to the momentum group $G$, which acts as translations under the particular isomorphism between $\mathbb X$ and $G$ on the configuration space, the Stone-von Neumann theorem holds.
\\
Having a particular measure $\mu_o$ on $\mathbb X/G$ at our disposal, we can consider the associated Schr\"odinger representation on $\mathcal H=L^2(\mathbb X,\mu_o\times \mu_H)$: In some cases, there are suitable semiclassical states, which one can obtain as ''Weyl-transforms of Dirac-distributions on the classical phase space''. Let us explain, what we mean with this:
\\
Given a set of points in the classical phase space $(x_o,p_o)\in \mathbb X\times\mathfrak g$ we have to first choose a vacuum state $\Omega$, which is exists, and is apparent if we started with a GNS-representation. Then we use the heuristic ansatz
\begin{equation}
  \psi_{x_o,p_o} : x \mapsto \biggl(\pi((x,g)\mapsto\delta(g,\exp(ip_o))\delta(x,x_o))\Omega\biggr)(x),
\end{equation}
where the Dirac $\delta$ distributions are obviously not continuous functions, such that we have to approximate them by series of continuous functions. Particularly, since we have a given measure $\mu$ on the configuration space and a Haar measure $\mu_H$ on the momentum group, we can use the Hilbert-spaces $\mathcal H$ and $L^2(G,\mu_H)$ and, provided they are separable, find a countable dense basis $\{c_i\in \mathcal H:i \in \mathbb N\}$ and $\{m_j \in L^2(G,\mu_H):j \in \mathbb N\}$ consisting of continuous functions on $\mathbb X$, $G$ respectively, which enables us to rewrite the $\delta$ as $\delta(x,x_o)=\lim_{N\rightarrow\infty}\delta_N(x,x_o)=\lim_{N\rightarrow \infty} \sum_{i=1}^N \overline{c_i(x)}c_i(x_o)$ and similarly on $G$. Thus, we arrive at
\begin{equation}
  \psi_{x_o,p_o} : x \mapsto \biggl(\lim_{N\rightarrow\infty}\pi((x,g) \mapsto \delta_N(g,\exp(tp_o))\delta_N(x,x_o) ) \Omega\biggr)(x).
\end{equation}
Provided, the limit is in the Hilbert-space, we have a suitable candidate for semiclassical states. We do not have a proof of general existence, however in those examples, that we considered so far it always turned out, that these limits existed in the Hilbert-space. Particularly using an ordinary quantum harmonic oscillator, one obtains the coherent states as semiclassical states using this prescription.

\section{Lesson From Groupoid Embeddings}

We saw in the previous section, that groupoids, and in particular transformation groupoids, act as mediating structures, that stand in between classical and quantum systems. On the one hand they can still be interpreted as classical spaces with a topology, on the other hand, they already contain the noncommutative convolution product of the associated quantum system. Our strategy will thus be as follows: We will embed a ''smaller'' groupoid into a ''larger'' and we will use a groupoid module on the larger groupoid to induce the smaller groupoid and its groupoid module by induction using the methods of Morita equivalence of groupoids. Then we apply this structure to observables, i.e. functions on the groupoid. This will reveal the structure needed for the general construction in the next section.

\subsection{Full Embeddings of Transformation Groupoids}

Let us look at embeddings of a small Poisson-system into a larger one by considering the configuration space $\mathbb X=\mathbb R^n$ of $n$ particles moving in one dimension. The associated momentum group $G$ is $\mathbb R^n$, which acts on $\mathbb X$ by translation. Let us now assume, that only some of the particles, say particles number $(1,...,m)$, are detectable with our measurement apparatus, such that the ''physically interesting'' observables only depend on the state of particles $(1,...,m)$. We can extract these observables form the full observable algebra using the pull-back under the standard embedding 
$$ \eta: (x_1,...,x_n,g_1,...,g_n) \mapsto (x_1,...,x_m,0,...,0,g_1,...g_m,0,...,0), $$
which maps any observable of the full system into an observable on the physically relevant system by evaluating it at position $0$ and exponetiated momentum $e^{i0p}=1$ for physically irrelevant observables. We see, that this embedding is {\bf admissible}, in the sense, that it is an embedding of the small transformation groupoid into the large transformation groupoid.
\\
If we only consider the embedding of the configuration space $ \eta|_{\mathbb X_o} : (x_1,...,x_m) \mapsto (x_1,...,x_m,0,...,0)$, then there are many admissible embeddings: Each extension of the configuration space embedding $\eta|_{\mathbb X_o}$, that maps an associated exponentiated momentum by identity map or into the neutral element of the momentum group defines a groupoid morphism. However, physically only those embeddings are interesting, that embed the momentum group of each ''relevant particle'' non trivially. So how can we identify these ''physical'' embeddings?
\\
Basically, we want to only consider those embeddings, for which all exponentiated momenta of the full theory, that close on the embedded configuration space, are embedded. This leads to the definition of a ''full transformation subgroupoid'':
\begin{defi}
  An embedding $\eta$ of a transformation groupoid $\mathcal G(\mathbb X_o,G_o)$ into a transformation groupoid $\mathcal G(\mathbb X,G)$ is called {\bf full}, iff all $g \in G$, for which $\alpha_g: \eta(\mathbb X_o) \rightarrow \eta(\mathbb X_o)$ closes (i.e. those elements that close on $\mathbb X_o$) are in the image $\eta(G_o)$ of $G_o$.
\end{defi}
The physical example illustrated, that full embeddings are the physically interesting ones. On the other hand, the restriction $\eta|_{\mathbb X_o}$ of a full embedding to the configuration space $\mathbb X_o$ contains already all the information of $\eta$:
\begin{lem}
  A full embedding between two free transformation groupoids is completely determined by its restriction to $\mathbb X_o$.
\end{lem}
proof: Due to the freeness of the group action of the full groupoid, there is a bijection between the groupoid elements $(x,g)$ and pairs of points $(x,y)$ in the same $G$-orbit, given by $(x,g) \leftrightarrow (x,y=\alpha_g (x))$. Fullness requires that all orbit pairs $(x,y)$, where $x$ and $y$ are in the image of $\eta(\mathbb X_o)$ are realized. Thus, by the freeness of the group action in the small groupoid, we have completely determined the embedding of the smaller groupoid. $\square$
\\
In turn, since the full embeddings are physically interesting, we can use this lemma and consider only the restriction $\eta|_{\mathbb X_o}$ of an embedding to the configuration space and calculate the full embedding. Since the configuration variables embed as a commutative subalgebra, we can use the pull-back trick for them. Thus our strategy will be to use the pull-back trick for this subalgebra and extend this construction to the rest of the algebra.
\\
There is a deeper mathematical reason for considering full embeddings only, which stems from Morita theory: Consider ordinary two-dimensional mechanics embedded into three-dimensional mechanics, using the constraint $x_3=0$. If we embedded the elements of the associated transformation groupoid by:
\begin{equation}
  \eta: (x_1,x_2,g_1,e) \mapsto (x_1,x_2,0,g_1,e,e),
\end{equation}
where $e$ denotes the unit element in the momentum group, then we can quantize both (1) three-dimensional quantum mechanics as the ''larger system'' and the pull-back under the embedding as the ''smaller system''. However, upon quantization, we see, that the Stone-von-Neumann theorem holds for the larger system, thus giving it only one character, corresponding to a zero-dimensional space, whereas the smaller system turns out to be Morita equivalent to $C(\mathbb R)$, whose spectrum is one-dimensional. In order to avoid, that the ''smaller system'' becomes in a sense larger upon quantization, we have to demand, that the ''structure of the noncommutativity'' is preserved under the embedding. For transformation group systems, this is done by restricting oneself to full embeddings.

\subsection{Morita Equivalence for Groupoids: Induced Groupoid Modules}

We encountered groupoids as a mathematical framework, that can be used to describe a semi-direct product, which we called transformation groupoid. However, a simple way of thinking about a groupoid is to view it as a generalization of a group with many units, which is $\mathcal G^o$. And just as groups can act as transformations on spaces, one can define the action of a groupoid $\mathcal G$ acting as transformations on a space $\mathbb X$. The difference is, that a groupoid has many units, thus we have to specify, which unit is the right identity element at a given point, which means we need in addition to an action $\mu: \mathcal G \times \mathbb X \rightarrow \mathbb X$ a {\bf momentum map} $\rho:\mathbb X \rightarrow \mathcal G^o$. The formal definition is then\cite{muhly}:
\\
Given a Groupoid $\mathcal G$ with unit set $\mathcal G_o$ and range and source maps $r,s$, then a momentum map $\rho: \mathbb X \rightarrow \mathcal G_o$ is a continuous, open map form a locally compact space $\mathbb X$ into the unit space, defining ''which unit is the appropriate for a given element $x \in \mathbb X$''. This defines the set of composable elements $\mathcal G * \mathbb X :=\{(g,x)\in \mathcal G \times \mathbb X: s(g)=\rho(x)\}$, which enables us to define an action $\mu$ of $\mathcal G$ on $\mathbb X$ as a map $\mu: \mathcal G * \mathbb X \rightarrow \mathbb X$, such that 
\begin{itemize}
  \item $\rho(\mu(g,x))=r(g)$ whenever $(g,x) \in \mathcal G * \mathbb X$
  \item $\mu(e(\rho(x)),x)=x$ $\forall x \in \mathbb X$, where $e(x)$ denotes the unit element at $x$
  \item $\mu(g_1g_2,x)=\mu(g_1,\mu(g_2,x))$ whenever $(g_1,g_2) \in G^2$ and $(g_2,x) \in \mathcal G * \mathbb X$ 
\end{itemize}
The last line defines a left action of $\mathcal G$, however defining the composition to be a right composition above, we can also define a right action of $\mathcal G$ on $\mathbb X$. We call a locally compact space $\mathbb X$ which carries a left $\mathcal G$-action $\mu$ a left $\mathcal G$-module and similarly a space that carries a right action of a groupoid $\mathcal H$ a right $\mathcal H$-module. If we have one space, that is both a left $\mathcal G$-module and a right $\mathcal H$-module, then we call it a $\mathcal G$-$\mathcal H$-bimodule, if the two actions commute, i.e.:
\begin{itemize}
  \item $\rho_G(x h)=\rho_G(x)$ for all $(x,h)\in \mathbb X * \mathcal H$ and $\rho_H(g x)=\rho_H(x)$ for all $(g,x) \in \mathcal G * \mathbb X$
  \item $g (x h) = (g x) h$ for all $(g,x) \in \mathcal G * \mathbb X$ and for all $(x,h) \in \mathbb X * \mathcal H$.
\end{itemize}
A $\mathcal G$-$\mathcal H$-bimodule $\mathbb X$ is called a Morita equivalence bimodule, if it satisfies:
\begin{itemize}
  \item The actions of $\mathcal G$ and of $\mathcal H$ are free.
  \item Both actions are proper, i.e. the map $(g,x)\mapsto (\mu(g,x),x)$ is proper for all $(g,x) \in \mathcal G * \mathbb X$ and similarly for $\mathcal H$.
  \item The momentum map $\rho_G:\mathbb X \rightarrow \mathcal G_o$ reduces to a bijection from $\mathbb X / \mathcal H$ to the unit space $\mathcal G_o$ and
  \item the momentum map $\rho_H:\mathbb X \rightarrow \mathcal G_o$ reduces to a bijection from $\mathbb X/\mathcal G$ to the unit space $\mathcal G_o$. 
\end{itemize}
If there exists a $\mathcal G$-$\mathcal H$-bimodule, then we call the groupoids Morita equivalent and it turns out that two Morita equivalent groupoids have equivalent categories of left modules, because we can use the $\mathcal G$-$\mathcal H$-equivalence bimodule $\mathbb X$ to induce a module on $\mathcal G$ given an $\mathcal H$-module $Y$ as $\mathbb X * \mathbb Y/\sim$, where $(xh,y)\sim(x,hy)$ and $\mathbb X * \mathbb Y =\{(x,y)\in \mathbb X \times \mathbb Y: \rho_X(x)=\rho_Y(y)\}$.
\\
More importantly, given a free and proper $\mathcal G$-module $\mathbb X$ with surjective momentum map $\rho$, we can construct a Morita equivalent groupoid from this data using the following steps:
\begin{enumerate}
  \item We start out with taking the double space $\mathbb X * \mathbb X:=\{(x,y)\in \mathbb X \times \mathbb X: \rho(x)=\rho(y)\}$.
  \item We define the diagonal action of $\mathcal G$ on the double space by
  $$ g \triangleright (x,y) := (g\triangleright x, g\triangleright y). $$
  \item The induced groupoid space is then given by $\mathcal H:=\mathbb X * \mathbb X / \mathcal G$, where $\mathcal G$ acts by the diagonal action.
  \item The induced unit space $\mathcal H_o$ is given by the $\mathcal G$-orbits in $\mathbb X$, together with the the induced source and range maps which take the left resp. right part of the double groupoid into the corresponding orbit.
  \item The groupoid composition law is induced through $$ [(x,y)]\circ [(y,z)]:= [(x,z)], $$
  where the square brackets denote the equivalence classes under the diagonal action of $\mathcal G$ on the double set $(x,y)\in\mathbb X*\mathbb X$.
  \item The induced action of the groupoid on the module $\mathbb X$ is defined through
   $$ x \triangleleft [(x,y)] := y, $$
   and the associated momentum map is given by taking $x$ into its $\mathcal G$ orbit.
\end{enumerate}
In order to introduce our construction for $C^*$-algebras, we will make intensive use of this construction. This is justified by an important theorem due to Muhly, Renault and Williams, which states, that given two locally compact Morita equivalent groupoids $\mathcal G$ and $\mathcal H$ and given a Haar system for each groupoid, then we the convolution $C^*$-algebras associated with the two groupoids are Morita equivalent as $C^*$-algebras.

\subsection{Canonical Groupoid-Module for Transformation Groupoids}

In the previous subsection, we have only abstractly mentioned groupoid modules. However for transformation groupoids $\mathcal G(\mathbb X,G)$, there is a canonical groupoid module $\mathbb X$, that corresponds to the canonical representation of the associated $C^*$-algebra $C^*(\mathbb X,G)$ on the Hilbert-space $L^2(\mathbb X,\mu)$. 
\\ 
Let us consider this module over $\mathcal G(\mathbb X,G)$ in more detail: It is given by the space $\mathbb X$ and the momentum map $\rho=id_{\mathbb X}$ is given by the identity in $\mathbb X$, since $\mathcal G^o=\mathbb X$. Thus, the set of composable elements is: 
\begin{equation} 
  \mathcal G * \mathbb X = \{(\gamma,x):\gamma \in \mathcal G, x \in \mathbb X: s(\gamma)=\rho(x) \}=\{((x,g),x): (x,g)\in\mathcal G, x \in 
\mathbb X\}. 
\end{equation} 
The action $\mu$ of $\mathcal G$ on $\mathbb X$ is then given in terms of the action $\alpha$ of the momentum group on the configuration space: 
\begin{equation} 
  \mu_{(x,g)}: x \mapsto \alpha_g(x). 
\end{equation} 
The freeness of $\alpha$ lets us identify $\gamma=(x,g)$ with $(x,y=\alpha_g(x))$, such that the action $\mu$ reduces to $\mu_{(x,y)}: x \mapsto y$. \\ Using this groupoid module, we can preform a warm-up exercise, that prepares for the strategy in the next subsections: Let us calculate the induced groupoid from the groupoid module $\mathbb X$. Following the construction in the previous section, we get: \begin{enumerate} 
  \item The underlying set is the double space: $\mathbb X * \mathbb X$, which is 
    $$ \mathbb X * \mathbb X =\{ (x,y):x,y \in \mathbb X, \rho(x)=\rho(y) \} =\{(x,x): x \in 
\mathbb X \} \sim \mathbb X, $$ 
    where we used the momentum map $\rho=id_{\mathbb X}$ and associated $(x,x)\sim (x)$. 
  \item The diagonal action of $\mathcal G(\mathbb X,G)$ on $\mathbb X * \mathbb X$ is 
  $$ (x,g) \triangleright (x,x) = (\alpha_g(x),\alpha_g(x)), $$ 
  such that $(x,g) \triangleright (x) = (\alpha_g(x))$. 
  \item The induced groupoid space is then given by the orbit space $\mathbb X * \mathbb X / \mathcal G$:  
    $$(x) \sim (y) \textrm{ iff } \exists \gamma \in \mathcal G : \gamma \triangleright (x) = (y), $$ 
    which implies that the elements of the induced groupoid are labelled by $G$-orbits $[x]_G$ in $\mathbb X$: 
    $$ \mathcal G_{ind}=\mathbb X * \mathbb X / \mathcal G =\mathbb X / G. $$ 
  \item The unit space of the induced groupoid is $\mathbb X / \mathcal G$, which is also $\mathbb X / G$.      The induced source and range maps are then: $s([x]_G)=r([x]_G)=[x]_G$. \item The groupoid composition law is trivial, since all composable pairs of elements act as unit elements $[x]_G \circ [x]_G =[x]_G$.   
  \item The action of the induced groupoid on the groupoid module module $\mathbb X$ is $$ x \triangleleft ([x]_G) = x, $$ and the momentum map is induced to be $\mu(x) = [x]_G$. 
\end{enumerate} 
We see, that this groupoid is the trivial groupoid over the space of $G$-orbits in $\mathbb X$, which is the reason, why we call it the orbit groupoid. Next, using the methods of Morita equivalence for groupoids, let us reconstruct the transformation groupoid from the induced action of the orbit groupoid on the groupoid module $\mathbb X$. 
\begin{enumerate} 
  \item The double space $\mathbb X * \mathbb X$ is 
$$ \mathbb X * \mathbb X = \{(x,y)\in \mathbb X \times \mathbb X: [x]_G=[y]_g\} 
$$ the set of all pairs of points on $\mathbb X$, that lie in the same $G$-
orbit. 
  \item The orbit groupoid acts trivially on the double space by:
    $$ [x]_G \triangleright (x,y) := (x,y) $$.
  \item Thus, the groupoid space $\mathbb X * \mathbb X / \mathcal G$ is given by the space of $G$-orbits: 
  $$ \mathcal G_{ind} = \{ (x,y) \in \mathbb X: \exists g \in G: y =\alpha_g(x) \}, $$
  which lets us associate $(x,y=\alpha_g(x))$ with $(x,g)$.
  \item The unit space is $\mathbb X / \mathcal G$, which coincides with $\mathbb X$ due to the trivial action of the orbit groupoid. The induced source and range maps are $s(x,y)=x$ and $r(x,y)=y$.
  \item The groupoid composition law is induced to be:
    $$ (x,y=\alpha_g(x)) \circ (y,z=\alpha_h(y)) = (x,z=\alpha_{hg} x). $$
  \item The induced momentum map is $\rho: \mathbb X \rightarrow \mathbb X = Id_{\mathbb X}$, which lets us calculate the induced action on $\mathbb X$ as:
  $$ \mu_{(x,y)} x = y \textrm{ or: } \mu_{(x,g)} x = \alpha_g(x). $$
\end{enumerate} 
Thus, we have reconstructed the transformation groupoid as the induced groupoid from the action of the orbit groupoid on the configuration space as prescribed by the induction of groupoids from Morita theory.
\\
Since the configuration space ''is commutative'' even in a quantum theory, our strategy for the next subsection is as follows: We have reconstructed the transformation groupoid from the orbit groupoid, which is a trivial groupoid, such that its algebra corresponds to a commutative $C^*$-algebra, which means that we can use the pull-back trick. This suggests, that we embed the a full subgroupoid and calculate its induced orbit groupoid and then reconstruct the reduced transformation groupoid from the induced structure. We will use this to observe the structure, that is needed to reconstruct the embedded groupoid. Thereafter we will use something similar to ''the pull-back'' under this structure on functions on the groupoid to reveal a general construction, that we can apply to quantum theories.

\subsection{Inducing an Embedded Subgroupoid}

Let us consider a full embedding $\eta$ of a transformation groupoid $\mathcal G_o(\mathbb X_o,G_o)$ into $\mathcal G(\mathbb X,G)$ given by its restriction $\eta|_{\mathbb X_o}$ to $\mathbb X_o$. Let us now do the following: First, we restrict the canonical action of the groupoid $\mathcal G$ on the canonical groupoid module $\mathbb X$ to $\eta(\mathcal G_o)$ to calculate the restricted orbit groupoid. Then we will use the action of the restricted orbit groupoid on the full configuration space $\mathbb X$ to calculate the reduced groupoid $\mathcal G_o$ using the induction of groupoids.
\\
Construction of the restricted orbit groupoid:
\begin{enumerate}
  \item The construction of the double set $\mathbb X * \mathbb X$ using the formula 
  $$ \mathbb X_o * \mathbb X_o =\{(x,y)\in \mathbb X \times \mathbb X: \rho_o(x)=\rho_o(y)\} $$
  has only a meaning for $x \in \mathbb X_o$. Thus we have to introduce the constraint 
  $$ \chi (x) = \biggl\{ \begin{array}{cl}
                           \emptyset &\textrm{ for } x \in \mathbb X \setminus \mathbb X_o\\
                           x & \textrm{ for } x \in \mathbb X_o
                         \end{array}\biggr. $$
  Hence, when constructing the double set, we have to consider the pair $(\mathbb X * \mathbb X, \chi)$, where $\mathbb X * \mathbb X =\{(x,y) \in \mathbb X_o \times \mathbb X_o: \exists g_o \in G_o : y =\alpha_g(x)\}$ is constructed from the momentum map of $\mathbb G_o$.
  \item The diagonal action of $\mathcal G_o$ on $\mathbb X * \mathbb X$ preserves the constraint $\chi$, such that we can use:
  $$ (x,g_o) \triangleright (x,y=\alpha_h(x)) = (\alpha_g(x),\alpha_{gh}(x)). $$
  \item Since the diagonal action preserved the constraint, we can calculate the restricted groupoid space again as $\mathbb X * \mathbb X / \mathcal G_o$, which yields:
  $$ \mathcal G_r =\{[x]_{G_o}: x \in \mathbb X_o\} $$
  the $G_o$-orbits in $\mathbb X$ by the canonical identification $([x]_{G_o},[x]_{G_o})\sim [x]_{G_o}$.
  \item The induced unit space of the restricted groupiod is the space of $G_o$-orbits in $\mathbb X_o$. The source and range maps are the just the identity maps as before $s([x]_{G_o})=[x]_{G_o}$ and $r([x]_{G_o})=[x]_{G_o}$.
  \item The induced composition law is trivial $[x]_{G_o} \circ [x]_{G_o} = [x]_{G_o}$.
  \item The action of the restricted groupoid on $\mathbb X$ is only well defined, when the constraint $\chi$ is not void, thus the momentum map is $\rho: \mathbb X_o \rightarrow \mathbb X_o/G_o: x_o \mapsto [x_o]_{g_o}$ and the action is induced to be:
  $$ \mu_{[x_o]_{G_o}} (x_o)= x_o$$
\end{enumerate}
Notice, that we need the constraint $\chi$ in addition to the full groupoid $\mathcal G$ and the restriction $\eta|_{\mathbb X_o}$ of the embedding of the reduced groupoid to be able to construct the restricted orbit groupoid. This additional constraint will play an important role in the definition of a quantum embedding in the next section.
\\
There is an interesting similarity to group averaging: In the group averaging approach, one needs a group $H$ with an action on $\mathbb X$, such that this action is compatible with the groupoid action. Furthermore, one needs that the $H$-orbits in $\mathbb X$ are labelled precisely by the points in $\mathbb X_o$, i.e. $\mathbb X_o = \mathbb X / H$. In this case we could have proceed as follows: we would have enlarged $\mathbb X * \mathbb X$ to pairs of elements of $\mathbb X$ in the same $G\times H$-orbits: $ \{ (x,y) \in \mathbb X \times \mathbb X: \exists g \in G, h \in H : y = \alpha_{g \otimes h}(x)\}$. We would have found he same orbit groupoid, but instead of the constraint $\chi$ we would have the $H$-orbits in the action of the restricted orbit groupoid.
\\
Now, that we have the restricted orbit groupoid and the constraint $\chi$ at our disposal, we are in the position to reconstruct the canonical action of $\mathbb G_o$ on $\mathbb X_o$ from the data $(\mathcal G, \mathbb X, \eta|_{\mathbb X_o}, \chi)$, where the first three where encountered in the construction of the restricted orbit groupoid, where the constraint appeared:
\begin{enumerate}
  \item The double set is constructed as always in the induction theory, with the exception that the constraint is taken care of by only considering pairs of points in the same orbit, for which $\chi$ is non-empty:
  $$ \mathbb X * \mathbb X :=\{ (x,y): \rho_o(x)=\rho_o(y) \textrm{ and } \chi(x)\ne \emptyset \ne \chi(y) \}, $$
  where we are again able to make the canonical association $(x_o,y_o=\alpha_{g_o}(x_o)) \leftrightarrow (x_o,g_o)$.
  \item The diagonal action of the restricted orbit groupoid on the double set is well defined, and again trivial $[x_o]_G \triangleright (x_o,y_o=\alpha_{g_o}(x_o)) = (x_o,y_o)$.
  \item Thus, the groupoid space $\mathbb X * \mathbb X / \mathcal G_r=\mathbb X * \mathbb X$.
  \item The base is constructed by restricting again $\mathbb X / \mathcal G_r$ to the points for which the constraint is nonempty:
  $$ \mathcal G_o = \{x \in \mathbb X/ \mathcal G_r: \chi(x) \ne \emptyset\}=\mathbb X_o. $$
  \item The induced groupoid composition law is simply: $(x_o,y_o) \circ (y_o,z_o)=(x_o,z_o)$, or equivalently $(x_o,g_o)\circ(\alpha_{g_o}(x_o),h_o) =(x_o,g_o h_o)$.
  \item Before we derive the induced action of the induced groupoid on $\chi(\mathbb X)=\mathbb X_o$, let us first calculate the momentum map $\rho: (\mathbb X_o \rightarrow \mathbb X_o / \mathcal G_r=\mathbb X_o)$, which is given  by $\rho = id_{\mathbb X_o}$. The induced action $\mu$ is then: $(x_o,y_o) \triangleright x_o \mapsto y_o$.
\end{enumerate}
Thus, using only a slight modification of the procedure of groupoid induction, we reconstructed the embedded groupoid from rather simple data, which only needed $\eta|_{\mathbb X_o}, \chi$ besides the full orbit groupoid. Obviously, due to the modification of the induction process, the reconstructed groupoid will in general not be Morita equivalent to the full groupoid, that we started out with.

\subsection{Reduced Sensitivity on a Transformation Groupoid}

A groupoid $\mathcal G$ is on the one hand a classical space which, in the case of a transformation groupoid, can be endowed with a topology induced from the configuration space and from the momentum group, provided the action is proper (and as we assume trough out this paper also free). As such, if both the configuration space $\mathbb X$ and the momentum group $G$ are locally compact, we have a canonical commutative $C^*$-algebra $C(\mathcal G)$ of continuous complex, that vanish at topological infinity, which represent the ''classical observables''. Since these classical observables commute, we can indeed use the pull-back trick, s.t. given an embedding $\eta$ of a ''small transformation groupoid'' $\mathcal G_o$ into a ''larger transformation groupoid'' $\mathcal G$ we can use
\begin{equation}
  \eta^*: C(\mathcal G) \rightarrow C(\mathcal G_o),
\end{equation}
to extract ''relevant degrees of freedom'' as in any classical theory. However, let us proceed differently here and consider full embeddings, such that it is sufficient to consider the restriction of $\eta|_{\mathbb X_o}$ to the configuration space, which will still be commutative after quantization, which at the level of observables consists mainly of replacing the commutative point wise product with the noncommutative convolution product.
\\
Using the point of view, that the sensitivity of our measurements is reduced to only resolve observables on the reduced groupoid, the pull-back under a full embedding means nothing else, than that the reduced sensitivity is already completely determined by a reduced sensitivity of the configuration variables. Thus, let us consider the canonical action of a function $F \in C(\mathcal G)$ on the transformation groupoid on a function $f \in C(\mathbb X)$ on the canonical groupoid module $\mathbb X$, which represents a configuration observable:
\begin{equation}
  F \triangleright f : x \mapsto \int_G d\mu_H(g) F(x,g) f(g^{-1} \triangleright x).
\end{equation}
In the next section we will construct the analogue to $\mathbb X * \mathbb X / \mathcal G$ for the configuration variables, which is in an obvious way given by the rank-one operators, that we used to construct Morita equivalence. But before we do this, let us consider, what structure we need to encode the pull-back under a full embedding using transformations $F \in C(\mathcal G)$, which act as linear transformations on the configuration variables:
\begin{enumerate}
  \item The transformations $F$ acting on any configuration variable $f \in C(\mathbb X)$ have to result in a configuration variable $f_o \in \mathbb C(\mathbb X_o)$ on the embedded configuration space $\mathbb X_o$. This can be achieved using a linear map $P$, which acts as the pull-back under the embedding $\eta$ on the transformation, i.e. $P(F \triangleright f) = \eta^*(F \triangleright f) \in C(\mathbb X_o)$. In other words, $P$ factors the ideal $\mathfrak I$ of functions out, that vanish at the closure of $\eta(\mathbb X_o)$: $\mathfrak I = \{f \in C(\mathbb X): f(x)=0 \forall x \in \eta(\mathbb X_o)\}$.
  \item As our construction shall result in transformations, that transform a configuration observable $f_o \in C(\mathbb X_o)$ on the embedded space into a variable on the embedded space, we need to ''inflate'' $f_o\in C(\mathbb X_o)$ to an element $f(f_o)$ of $C(\mathbb X)$, such that it can be plugged into $P(F\triangleright f(f_o))$. This is the point, where the constraint becomes important: Basically we search for continuous functions, that are something like the ''the inverse of $P$'', which we denote by $i: C(\mathbb X_o) \rightarrow C(\mathbb X)$. Obviously, since $i$ has to invert the constraint $\chi$ for the momentum map, it is not unique. But on the other hand if we satisfy the constraint that $P \circ i = id _{img(P)}$ and $i \circ P = id_{img(i)}$, then the formula $P(F \triangleright i(v))$ defines a set of linear transformations on $C(\mathbb X_o)$.
\end{enumerate}
Obviously, the choice of $i$ is not unique, since it is physically speaking the ''choice of a gauge'', if we consider $\chi$ as a true constraint. Although not unique, given any $P$, there always exists an $i$ by the axiom of choice. Putting these two ingredients together, we obtain the definition of a quantum Poisson embedding:
\begin{defi}
  Given an embedding $\eta$ of a space $\mathbb X_o \rightarrow \mathbb X$, we call a pair $(P,i)$ of linear maps between $C(\mathbb X)$ and $C(\mathbb X_o)$, where $P=\eta^*$, a {\bf quantum Poisson embedding}, iff $P\circ i=id_{C(\mathbb X_o)}$ and $i \circ P=id_{img(i)}$.
\end{defi}
Given a pair $(P,i)$, we can reconstruct the corresponding full embedding $\eta$ using the correspondence of maximal ideals of a commutative algebra and the points of the spectrum of the algebra: First, we can associate $(x,g)$ with pairs of points in $\mathbb X$: $(x,y=\alpha_g(x))$ as before. Then using the correspondence of points $x\in \mathbb X$ with maximal ideals $\mathfrak I_x = ker(\chi_x)$, where $\chi_x$ is the character corresponding to the evaluation of a function at $x$, we can associate pairs of maximal ideals $(x,g) \leftrightarrow (\mathfrak I_x,\mathfrak I_y)$. Then we use $P$, which maps a maximal ideal $\mathfrak I_x$ only to a maximal ideal of $C(\mathbb X_o)$, if $x \in \mathbb X_o$. Using only those pairs, that correspond to pairs $(x_o,y_o)$, where $x_o,y_o \in \mathbb X_o$, we obtain the desired embedding. On the other hand starting out with a pair of maximal ideals $(\mathfrak I_{x_o},\mathfrak I_{y_o})$ in $C(\mathbb X_o)$, we can use $i$ and find the maximal ideal in $C(X)$, such that (1) its $P$-image is a maximal ideal in $C(\mathbb X_o)$ and (2) contains all elements of $i(\mathfrak I_{x_o})$ resp. $i(\mathfrak I_{y_o})$.

\section{General Construction}

We have learned through the exercise of reconstructing an embedded transformation groupoid form the restriction of an orbit groupoid, that a fully embedded groupoid can be reconstructed from the canonical module for the orbit groupoid together with the restriction of the embedding $\eta|_{\mathbb X_o}$ and the constraint $\chi$, which restricts the momentum map to the embedded configuration space. In terms of observables on the groupoid, we saw, that the corresponding structures can be provided by a quantum Poisson embedding, consisting of a compatible pair of linear maps $(P,i)$ between the functions $C(\mathbb X)$ on the canonical module and the functions on $\chi$-image of the canonical module $C(\mathbb X_o)$. 
\\
The link between the induction of groupoids using groupoid modules and induction of $C^*$-algebras using $C^*$-modules\footnote{This link is rather strong: Given the fact that two groupoids are Morita equivalent, then it follows that the associated $C^*$-algebras are Morita equivalent, too\cite{muhly}.} suggests, that we use this strategy to induce a ''fully Poisson-embedded quantum algebra'' from a given quantum algebra. Before we carry out the explicit construction, let us consider a simple example in which we illustrate how the modifications that we have used in groupoid induction to reconstruct the embedded groupoid can be translated into the induction process for $C^*$-algebras.

\subsection{Introductory Example}

The possibly simplest example of quantum mechanics is quantum mechanics on a circle. The goal of this section is to extract quantum mechanics on the circle from quantum mechanics on a 2-torus using reduced sensitivity: Say, the torus is coordinized by $(\phi_1,\phi_2): \phi_i \in [\,0,2 \pi)$, then we can assume, that our measurements only resolve the coordinate $\phi_1$ and are insensitive to $\phi_2$ for some reason, which results that the measurements at our disposal only resolve quantum mechanics on a circle.
\\
The example in this section is chosen in such a way, that the extraction of quantum mechanics on the circle from quantum mechanics on the torus becomes nothing else but ''neglecting the $\phi_2$-dependence'' in the Schr\"odinger representation of the theory. However, at the end of this section, we will have disguised this fact in such a way, that $\phi_2$ does not appear in our construction anymore, but the quantum Poisson-embedding $(P,i)$ is put to the foreground.
\\
Let us start by describing the quantum algebras and their canonical modules:
\\
The configuration variables of quantum mechanics on a torus are continuous complex-valued functions on the torus, which we view as the direct product of two copies of $U(1)$, i.e. we consider $C(U(1)^2)$ as the algebra of configuration variables. The group $W(\lambda_1,\lambda_2)\in U(1)^2$ of exponentiated momenta  acts ''unitarily'' as pull-backs under translations on $C(U(1)^2)$, satisfying the Heisenberg commutation relations with configuration variables $f\in C(U(1)^2)$:
\begin{equation}
  W(\lambda_1,\lambda_2)^* f W(\lambda_1,\lambda_2): (\theta_1,\theta_2) \mapsto f(\theta_1-\lambda_1 \textrm{ mod } 2 \pi, \theta_2 - \lambda_2 \textrm{ mod }2 \pi).
\end{equation}
Having the elementary configuration and momentum operators at our disposal, we can use Weyl-quantization to associate a quantum observable $\hat F$ to any function $F$ on $U(1)^2 \times U(1)^2$, where the first $U(1)^2$ denotes the configuration space and the second the momentum group, by setting:
\begin{equation}
  \hat F := \int_{U(1)^2} d\lambda_1 d\lambda_2 F(.,.,\lambda_1,\lambda_2) W(\lambda_1,\lambda_2),
\end{equation}
where $F(.,.,\lambda_1,\lambda_2)$ is understood as a function $f(.,.)$ on the configuration space. This induces the operator product of two quantum variables $\hat F_1,\hat F_2$. The canonical representation on the module $C(U(1)^2)$ (given by the configuration variables) is:
\begin{equation}
  \hat F v : (\phi_1,\phi_2) \mapsto \int_{U(1)^2} d\lambda_1 d\lambda_2 F(\phi_1,\phi_2,\lambda_1,\lambda_2) v(\phi_1-\lambda_1\textrm{ mod }2 \pi,\phi_2-\lambda_2\textrm{ mod }2 \pi).
\end{equation}
As we have done in quantum mechanics on the real line, we consider the covariant pair of representations consisting of configuration variables and the momentum group, which turns out to be rather practical. Their action on an element $v \in C(U(1)^2)$ of the canonical module is:
\begin{equation}
  \begin{array}{rcl}
    \hat f v : &(\phi_1,\phi_2)  \mapsto  & f(\phi_1,\phi_2) v(\phi_1,\phi_2)\\
    \hat W(\lambda_1,\lambda_2) v : & (\phi_1,\phi_2) \mapsto & v(\phi_1-\lambda_1\textrm{ mod }2 \pi,\phi_2-\lambda_2\textrm{ mod }2 \pi).
  \end{array}
\end{equation}
Quantum mechanics on the circle is then simply the restriction of these variables to one copy of $U(1)$ as configuration space and one copy as momentum group. The corresponding configuration operators $\hat f$ and momentum group elements $\hat W$ as well as composite Weyl-operators $\hat F$ act on elements $v \in C(U(1))$ of the canonical module by:
\begin{equation}
  \begin{array}{rcl}
    \hat f v : &(\phi)  \mapsto  & f(\phi) v(\phi)\\
    \hat W(\lambda) v : & (\phi) \mapsto & v(\phi-\lambda \textrm{ mod } 2 \pi)\\
   \hat F v : &(\phi) \mapsto & \int_{U(1)} d\lambda F(\phi,\lambda) v(\phi-\lambda \textrm{ mod }2 \pi).
  \end{array}
\end{equation}
Let us now bring the transformation groupoids into play: The transformation groupoid $\mathcal G$ of the torus system is given by elements of the form $(\phi_1,\phi_2,\lambda_1,\lambda_2)$ with the source and range maps:
\begin{equation}
  \begin{array}{rl}
    s(\phi_1,\phi_2,\lambda_1,\lambda_2)& = (\phi_1,\phi_2)\\
    r(\phi_1,\phi_2,\lambda_1,\lambda_2)& = (\phi_1-\lambda_1 \textrm{ mod }2 \pi,\phi_2-\lambda_2 \textrm{ mod }2 \pi)
  \end{array}
\end{equation}
The composition law is simply $(\phi_1,\phi_2,\lambda_1,\lambda_2)\circ(\phi_1-\lambda_1,\phi_2-\lambda_2,\lambda_1^\prime,\lambda_2^\prime)=(\phi_1,\phi_2,\lambda_1+\lambda_1^\prime,\lambda_2+\lambda_2^\prime)$. The action of the groupoid on the configuration space is $(\phi_1,\phi_2,\lambda_1,\lambda_2)\triangleright (\phi_1,\phi_2) = (\phi_1-\lambda_1 \textrm{ mod }2 \pi,\phi_2-\lambda_2 \textrm{ mod }2 \pi)$, where we used the momentum map $\rho=Id_{U(1)^2}$. The groupoid $\mathcal G_o$ associated with the circle system is obtained by simply deleting the entries with index $\,_2$ in our notation and the action of the circle groupoid on $U(1)$ is obtained in the same way. Using these two groupoids, the reduced sensitivity statement, that our measurements are only sensitive to the $\phi_1$-component, can be stated in terms of the groupoid map $\eta: \mathcal G_o \rightarrow \mathcal G$:
\begin{equation}
  \eta: (\phi,\lambda) \mapsto (\phi,0,\lambda,0).
\end{equation}
Let us now construct the associated quantum Poisson map, that corresponds to the pull-back under $\eta$. First, let us consider the restriction of $\eta$ to the configuration space $\mathbb X_o=U(1)$ and construct the map $P=(\eta|_{\mathbb X_o})^*$ by its action on an element $v$ of the canonical configuration module $C(U(1)^2)$ on the torus:
\begin{equation}
  P v : \phi \mapsto v(\phi,0).
\end{equation}
Secondly, considering the action of the embedding of the groupoid $\eta(\mathcal G_o)$ on the configuration space $U(1)^2$, we have the constraint for the momentum map, that restricts the domain of the momentum map to the embedding of the circle into the torus. In our case this is:
$$ \chi(\phi_1,\phi_2)=\biggl\{\begin{array}{l} \emptyset \textrm{ for } \phi_2 \ne 0 \\ (\phi_1,0) \textrm{ for } \phi_2 =0 \end{array} \biggr. $$
The map $i: C(U(1)) \rightarrow C(U(1)^2)$ has to ''pick a representative function'' on the torus for each function on the circle, such that it coincides with the original function at the embedding. An admissible map $i$ is:
\begin{equation}
  i v : (\phi) \mapsto v(\phi_1,\phi_2)|_{\chi \ne \emptyset} = v(\phi,0).
\end{equation}
Having the quantum Poisson embedding at our disposal and using the rank-one operators on the torus:
\begin{equation}
  \hat O_{f_1,f_2} = \hat f_1(\phi_1,\phi_2) \int_{U(1)^2}d \lambda_1 d \lambda_2   \biggl(\hat f_2(\phi_1-\lambda_1 \textrm{ mod }2 \pi,\phi_2-\lambda_2\textrm{ mod }2 \pi )\biggr) \hat W(\lambda_1,\lambda_2),
\end{equation}
we are in the position to define ''rank-one operators'' $O^\eta$ on the module of configuration variables on the circle $v \in C(U(1))$, given a pair $f_1,f_2\in C(U(1)^2)$ of elements of the canonical configuration module on the torus, by
\begin{equation}
  O^\eta_{f_1,f_2}: v \mapsto P(O_{f_1,f_2} i(v)),
\end{equation}
which can be written using more explicitly as:
\begin{equation}
  O^\eta_{f_1,f_2} v: \phi \mapsto f_1(\phi,0) \int_{U(1)^2} d\lambda_1 d\lambda_2 f_2(\phi-\lambda_1\textrm{ mod }2 \pi,2 \pi -\lambda_2) v(\phi-\lambda_1).
\end{equation}
Although this looks merely like we have tried to disguise the fact, that we ''simply neglected the $\phi_2$-dependence in the Schr\"odinger representation, to construct quantum mechanics on a circle as a subsystem of quantum mechanics on the torus, we have really gone the essential step forward: We have used a quantum Poisson embedding to calculate the quantum algebra of the circle system from the full quantum algebra. This puts us in the position to use the analogue to Rieffel induction, to obtain not only the quantum algebra, but also an induced representation for it.
\\
Let us consider the canonical Hilbert-space representation of quantum mechanics on the torus, which is given by the completion of the canonical module $C(U(1)^2)$ in the inner product $\langle f_1,f_2 \rangle := \int_{U(1)^2} d\phi_1 d\phi_2 \overline{f_1(\phi_1,\phi_2)} f_2(\phi_1,\phi_2)$, which yields the Hilbert-space of square integrable functions on $U(1)^2$. The trick, that we used earlier was to induce a second set of rank-one operators $T_{f_1,f_2}$, which was (1) determined by its action on the canonical module through $T_{f_2,f}: f_1 \mapsto O_{f_1,f_2} f$ and (2) took its values in the induced algebra, such that we where able to induce a state $\Omega$ on the induced algebra through a state $\omega$ on the original algebra by setting $\Omega(O_{f,g}):=\omega(T_{f,g})$. 
\\
We constructed the operators $O^\eta$ in such a way, that they where transformations on the circle module, such that we have to extend the condition (1) that determines the $T^\eta$ so they are transformations in the torus system (so the condition (2) can be satisfied). We can try to construct $V^\eta$ by using a combination of $P,i$:
\begin{equation}
  V^\eta_{f,g} : h \mapsto i(O^\eta_{h,f} P(g)),
\end{equation}
where $f,g,h \in C(U(1)^2)$. The operators $V^\eta$ are no longer rank-one operators, as we can see by calculating the action of $V^\eta$ any function of the kind $f=\sum_{m,n} A_{mn} e^{i (n \phi_1 + m \phi_2)}$ and compare it with the action on $f = \sum_n A_{0,n} e^{in \phi_1}$, which turn out to coincide. Although, we could make sense out of this by extending our definition of a state to the linear combinations, it is better to consider real rank-one operators $T^\eta_{f,g}$ associated with each $O^\eta_{f,g}$ by:
\begin{equation}
  T^\eta_{f,g} := O_{i(P(g)),i(P(f))},
\end{equation}
where $O_{f,g}$ is the above defined rank-one operator, which is clearly an element of the Weyl-algebra on the torus. Using these operators, we can again use the trick above and induce a state $\Omega$ on the span of the operators $O^\eta$ given a state on the algebra, that contains the operators $T^\eta$ by
\begin{equation}
  \Omega(O^\eta_{f,g}):=\omega(T^\eta_{f,g}).
\end{equation}
Let us now work these formulas out for the Schr\"odinger vacuum state $\omega$ on the torus, which is given by $\omega(\hat F)= \frac 1 {(2 \pi)^2}\int_{U(1)^2} d\phi_1 d\phi_2 F(\phi_1,\phi_2,0,0)$:
\begin{equation}
  \begin{array}{rcl}
  \Omega(O^\eta_{f,g})=\omega(T^\eta_{f,g})&=&\frac 1 {(2 \pi)^2}\int_{U(1)^2} d\phi_1 d\phi_2 f(\phi_1,0) \overline{g(\phi_1,0)}\\&=&\frac 1 {2 \pi}\int_{U(1)} d\phi f(\phi,0) \overline{g(\phi,0)},
  \end{array}
\end{equation}
which is the Schr\"odinger vacuum state on the circle. Again, we have just disguised the fact, that we extracted quantum mechanics on a circle from quantum mechanics on the torus by neglecting the ''$\phi_2$-dependence'' both in the observable algebra and in the Schr\"odinger representation. However, the essential step is again, that this procedure was done using the quantum Poisson embedding and that we did not have to refer to the ''$\phi_2$ dependence'' in our construction at the abstract level.

\subsection{Strategy}

The strategy for the next two subsections is the application of what we have done in the example of the previous section in the context of (1) a general transformation group system and (2) a general representation of the corresponding quantum algebra.
\\
The generalization to a general transformation group system is straight forward: Given a transformation group system based on the configuration space $\mathbb X$ and the group of exponentiated momenta $G$, we have already encountered the associated quantum algebra $C^*(\mathbb X,G)$ and its canonical module $C_c(\mathbb X)$. We will again use the formula for the rank-one operators $O_{f,g}$ for this $C^*$-algebra and decorate it with the quantum Poisson map $(P,i)$, such that we obtain the reduced quantum algebra as the span of the rank-one operators, that are defined through their action on a canonical module $C_c(\mathbb X_o)$ of functions on the embedded configuration space $\mathbb X_o$.
\\
For the induced representations, we will first proceed as we have done here and associate rank-one operators $T^\eta_{f,g}\in C(\mathbb X,G)$ to the operators $O^\eta_{f,g}$ again by decorating them with the quantum Poisson embedding $(P,i)$. This will allow us to use the same trick to induce a state $\Omega$ by setting $\Omega(O^\eta_{f,g}):=\omega(T^\eta_{f,g})$. Then we will use the fact, that any representation is a direct sum of GNS-representations out of particular vacuum states $\omega$ and use the induction of states to construct the induction of representations.\footnote{The induction of representations rather than states is the way, in which Rieffel induction is usually presented.} By constructing Rieffel induction for representations out of Rieffel induction for states, we will be able to ''read the generalization to our embedding problem off'' and obtain a construction for the induction of a representation of the embedded system.
\\
In the last subsection, we will then use the approximate identity, that we already encountered in section four to explore some of the properties of our construction. In particular, we will show, that we satisfy equ. (\ref{l1}) (the induction of the right observable algebra) and equ. (\ref{l2}) (the induction of the right state).

\subsection{Reduced Algebra}

Before we apply the embedding procedure to a general transformation group system, we will first describe the canonical $C^*$-module and apply the construction of the quantum Poisson map $(P,i)$ to this module.
\\
Given a transformation group $C^*$-algebra $C^*(\mathbb X,G)$, where the action of the group $G$ on the space $\mathbb X$ is assumed to be free and proper, we can construct the canonical $C^*$-module $C_c(\mathbb X)$, which can be endowed with the rank-one operators that are associated to the function:
\begin{equation}
  O_{f_1,f_2} : (x,g) \mapsto \Delta^{-\frac 1 2}(g) f_1(x) \overline{f_2(g^{-1} \triangleright x)},
\end{equation}
on the transformation groupoid, which is associated to two functions $f_1,f_2 \in C_c(\mathbb X)$. $\Delta$ denotes the modular function on the group. These rank one operators act on $f \in C(\mathbb X)$ as linear transformations by:
\begin{equation}
  O_{f_1,f_2} f : x \mapsto f_1(x) \int_G d\mu_H(g) \Delta^{- \frac 1 2}(g) \overline{f_2(g^{-1}\triangleright x)} f(g^{-1}\triangleright x).
\end{equation}
Now, we insert the quantum Poisson map, as we have done it above, to ensure, that it defines linear transformations on the canonical module $C(\mathbb X_o)$ of the embedded quantum algebra, which we define by the action on $h =P(f) \in C_c(\mathbb X_o)$:
\begin{equation}
  O^\eta_{f_1,f_2} h \mapsto P\biggl( x \mapsto f_1(x) \int_G d\mu_H(g) \Delta^{- \frac 1 2}(g) \overline{f_2(g^{-1}\triangleright x)} i(f)(g^{-1}\triangleright x) \biggr),
\end{equation}
where we used the abstract concept of defining a linear transformation in $C(\mathbb X_o)$ by setting $O^\eta : h \mapsto P(O i(h))$. We will later show, by using an approximate identity as in section four, that the algebra spanned by these rank-one operators is indeed dense in the transformation group algebra $C^*(\mathbb X_o,G_o)$ of the associated reduced system, that was embedded by $\eta$.
\\
Before we are able to calculate the induced representation in the next section, we need to construct associated the second set of rank-one operators $T^\eta_{f_1,f_2}$ to each $O^\eta_{f_1,f_2}$. In the previous example, we have just set $T^\eta_{f_1,f_2}: f \mapsto O_{i(P(f_1)),i(P(f_2))} P(f)$ with the justification, that it eliminated the $\phi_2$-dependence of the observables on the torus and resulted in quantum mechanics on the circle. 
\\
Let us now proceed closer to our induction process on groupoids: First, we notice that the formula $O_{f_1,f_2}f_3 = f_1 U_{f_2,f_3}$ induces, that the Morita induced algebra $C^*(C^*(\mathbb X,G),C(\mathbb X))$ is actually $C(\mathbb X/ G)$, by identifying the rank-one operators $U$ as:
\begin{equation}
  U_{f_1,f_2} f : [x]_G \mapsto f([x]_G) \biggl( [x]_G \mapsto \int_G d\mu_H(g) \overline{f_1(g^{-1}\triangleright x)}f_2(g^{-1}\triangleright x) \biggr),
\end{equation}
by identifying the group-averaged functions with functions on the group orbits under the big bracket. If we use the same induction process with $O^\eta$ on $P(f)=h \in C_c(\mathbb X_o)$, we obtain the induced rank-one operators the $U^\eta_{P(f_1),P(f_2)}$.
\\
Second, let us induce the operators $O_{f_1,f_2}$ again by imposing $O_{f_1,f_2}f_3 = f_1 U_{f_2,f_3}$, which indeed induces the operators $O_{f_1,f_2}$. Let us now repeat this induction step for $U^\eta_{P(f_1),P(f_2)}$ and consider $i$ as the inversion of the groupoid constraint $\chi$. This means that we have to consider the rank-one-operators as operators on the full module, which we can do by viewing then as operators on the image of $i$. Thus, we obtain by considering them as operators on the image of $i$, that the operators that are induced by $U^\eta$ are:
\begin{equation}
  T^\eta_{f_1,f_2}: P(f) \mapsto P(O_{i(P(f_1)),i(P(f_2))} P(f)),
\end{equation}
where $f,f_1,f_2 \in C_c(\mathbb X)$, which justifies the previously heuristic formula. Furthermore we see that the operators $T^\eta$ and $O^\eta$ are only compatible, when we restrict the action of the operators $O^\eta$ to the image of $i$. The mathematical formulation of noncommutative embeddings that use the vector bundle approach will require such a property.

\subsection{Induced Representation}

In the previous example we have induced a vacuum state for the embedded quantum algebra from a given vacuum state of the full quantum algebra. Using the fact, that a representation can be decomposed into cyclic representations, which we can construct as GNS representations out of a vacuum state, we can go ahead and generalize our construction to representations $(\mathfrak A, \pi, \mathcal H)$ of a $C^*$-algebra $\mathfrak A$ on a Hilbert-space $\mathcal H$. 
\\
Let us first consider a cyclic representation out of the cyclic representation out of the cyclic vector $\psi_1$, such that the set $\pi(a) \psi_1$ lies dense in the Hilbert space, such that the relation to the vacuum state $\omega$ is
\begin{equation}
  \omega(a) := \langle \psi_1, \pi(a) \psi_1 \rangle.
\end{equation}
This gives a in particular a representation of the rank-one operators $O^\eta_{f,g}$, which we use to define the induced vacuum state $\Omega$ by setting $\Omega(T^\eta_{f,g}):=\omega(O^\eta_{f,g})$:
\begin{equation}
  \Omega(T^\eta_{f,g}):=\omega(O^\eta_{i(P(f)),i(P(g))})=\langle \psi_1, \pi(O^\eta_{i(P(f)),i(P(g))}) \psi_2 \rangle.
\end{equation}
Using this induced state, we can carry out a GNS construction for the span of the rank-one operators $T^\eta$, which we used to define the reduced quantum algebra:
\\
To see, that $\Omega$ is a state, we first verify that it is linear by construction. Second, we verify that $O^\eta_{f,g}\ge 0$ implies that $T^\eta_{f,g}\ge 0$, which can be shown by noticing that $O^\eta_{f,g}$ is self-adjoined if and only if $f=g$, for which $O^\eta_{i(P(f)),i(P(f))}$ is clearly a non-negative operator. If we now use the fact that the action of the embedded algebra is faithful as linear transformations on the image of $i$, and if we furthermore realize that any positive element of the algebra of these linear transformations is a completion of a series of positive rank-one operators, then we conclude that the positive elements of the embedded algebra arise as series of positive rank-one operators. Using the linearity of $\omega$, we get the positivity of $\Omega$. A formal proof of this argument relies on an approximate identify that is the sum of positive rank-one operators $id=\sum_i T_{f_i,f_i}$, which we introduce in the next subsection, and the fact that $\sum_i T_{a f_i,a f_i}= \sum_i a T_{f_i,f_i} a^* \rightarrow a a^*$.
\\
Using the state $\Omega$ we calculate the Gelfand ideal, i.e. the null-space of the state $\Omega$ by:
\begin{equation}
  \mathfrak I := \{f\in C_c(\mathbb X): \Omega(T^\eta_{f,f})=0\}.
\end{equation}
Second, we define the canonical projection $\Pi$ from the span of the $T^\eta$, which we denote by $\mathfrak A_o$ to the quotient $\mathfrak A_o/ \mathfrak I$. Using this projection, we can define an inner product on $\mathfrak A_o / \mathfrak I$ by setting for $f,g \in C_c(\mathbb X)$:
\begin{equation}
  \langle \Pi(f),\Pi(g) \rangle := \Omega(T^\eta_{f,g}),
\end{equation}
which is well defined. Completing $\mathfrak A_o/ \mathfrak I$ in this inner product yields an induced Hilbert space $\mathcal H_{ind}$, which carries a natural representation of $\mathfrak A_o$ by setting:
\begin{equation}
  \pi(a) \Pi(\psi) := \Pi(\pi(a)\psi),
\end{equation}
which turns out to be well defined, since $\psi$ is in the closure of $\mathfrak A_o/\mathfrak I$. 
\\
The last step of inducing a representation consists of building an arbitrary representation of the full algebra $\mathfrak A$ as a direct sum of cyclic representations based on states $\omega_i$, carrying this construction out and then taking a direct sum of the induced representations based on the respective induced states $\Omega_i$.

\subsection{Properties}

Until now we have only spoken about the induced algebra $\mathfrak A_o$ as the span of the rank-one operators $T^\eta_{f,g}$. Let us now consider what this algebra is: We have claimed so far, that this is the fully embedded transformation group algebra $C(\mathbb X_o,G_o)$. The trick to prove this claim is to construct an appropriate approximate identity in $C(\mathbb X_o,G_o)$ similar to the example in section four.
\\
Let us first collect some preparations so we can formulate the requirements for the approximate identity in the case of a transformation group system:
\\
We have assumed that the embedded configuration space $\mathbb X_o$ as well as the full configuration space $\mathbb X$ are locally compact, such that the continuous functions of compact support in the respective spaces lie dense in the respective $C^*$-algebra of configuration variables, which consist of functions on the configuration space, that vanish at topological infinity. In turn, we can approximate any configuration variable by its restriction to larger and larger compact subsets of $\mathbb X$. Thus, we can again label an approximate identity by compact subsets.
\\
Moreover, we assumed that the action of the momentum group $G_o$ on the configuration space $\mathbb X_o$ is free and proper. This in turn implies that we can restrict arbitrarily small neighborhoods of the identity element in the momentum group by restricting oneself to those group elements, that transform at least one point in a given ''small'' compact set in the configuration space into another point in this set. It follows that we can impose the same conditions as in section four for the approximate identity and that we can construct it in a similar way:
\begin{equation}
  \begin{array}{rccl}
    id_{C,U(e),\epsilon}(x,g) & = & 0 & \textrm{for } g \textrm{ outside } U(e)\\
    |id_{C,U(e),\epsilon}(x,g)-1| & < & \epsilon & \textrm{for } x \in C,
  \end{array}
\end{equation}
where $C$ is a compact subset of the embedded configuration space $\mathbb X_o$, $U(e)$ is a neighbourhood of the identity element in the embedded momentum group $G_o$ and $\epsilon \in \mathbb R^+$. If we specify a net of compact subsets $C$, that approximates $\mathbb X_o$ as well as a net of neighborhoods $U(e)$ of $G_o$ that approximates the unit element, then the set $id_{C,U(e),\epsilon}$ defines an approximate identity in the embedded $C^*$-algebra. 
\\
The proof that the rank-one operators $T_{f,g}$ are really dense in the algebra $C^*(\mathbb X_o,G_o)$ is then verified by directly constructing the approximate identity as a sum of these rank-one operators. For a given triple $(C,U(e),\epsilon)$ this is again done by taking an open covering of the compact set $C$ by sets $U_i$ such that elements outside $U(e)$ have no point in any $U_i$, that is transformed into the same $U_i$. Then we can find continuous regularizations $\chi^\epsilon(U_i)$ of the characteristic functions of the open sets $U_i$, such that $|\sum_i T_{i(\chi^\epsilon(U_i)),i(\chi^\epsilon(U_i))} -1 | < \epsilon$ inside the set $C$, which implies, that $\sum_i T_{i(\chi^\epsilon(U_i)),i(\chi^\epsilon(U_i))}$ satisfies both conditions for the approximate identity.
\\
Realizing, that any element $a_o$ of the embedded algebra $C^*(\mathbb X_o,G_o)$ can be approximated by a sequence of the kind $\sum_i T_{i(a_o \chi^\epsilon(U_i)),i(\chi^\epsilon(U_i))} \rightarrow a_o$ we have shown, that our construction fulfilled the condition to reproduce the correct classical limit, since we have a unique way of going from $C^*(\mathbb X_o,G_o)$ to the associated Poisson system $P(\mathbb X_o,G_o)$.
\\
The second condition that we imposed on our construction was to reproduce the correct vacuum vector (or equivalently a set of correct vacuum expectation values). In terms of the rank one operators $O^\eta_{f,g}$ of the embedded quantum system, we have an apparent map $\mathcal E$ defined by
\begin{equation}
  \mathcal E: O^\eta_{f,g} \mapsto T^\eta_{f,g},
\end{equation}
which, by construction, is linear and dense in the embedded algebra. On the other hand, we constructed the induced state $\Omega$ exactly by setting $\Omega(O^\eta_{f,g}):=\omega(T^\eta_{f,g})$, such that there is an obvious matching of the induced vacuum state by simply inserting $\mathcal E$:
\begin{equation}
  \Omega(O^\eta_{f,g})=\omega(\mathcal E(O^\eta_{f,g})),
\end{equation}
which can be extended from the rank one operators to the induced algebra. Similarly, going through a GNS-construction, we obtain an obvious matching of matrix elements, which stems from the matching $\mathcal E$ and the canonical projection $\Pi: \mathfrak A_o / \mathfrak I$, where $\mathfrak I$ denotes the Gelfand ideal of the induced vacuum state $\Omega$.
\\
Another consequence of the existence of an approximate identity of the form described above is that we can use this approximate identity to ''forward'' operators from the full system to the reduced system. Using $id=\lim \sum_i T_{f_i,f_i}$his makes use of:
\begin{equation}
  id_o A = \lim \sum_i O_{f_i,f_i} A = \lim \sum_i O_{f_i, A^* f_i},
\end{equation}
which we can interpret as the ''relevant sector'' of the operator $A$, such that we obtain the relevant part by association:
\begin{equation}
  A_{red}:= \lim \sum_i T_{f_i,A^* f_i}.
\end{equation}
The interpretation of this operator is however different form the full system: It only contains the dependence of a measurement on the ''reduced system'', thus we have to interpret the forwarded operator $A_{red}$ not as the measurement of a particular observable in the full system, since generally many observables of the full system will coincide with an observable in the reduced system. This becomes clearer when we put this line of reasoning on its head: Given any observable in the full theory, there is an equivalence class of observables in the full system, that give us the same information about the reduced system. Physically, this is nothing else than to say, that there is more than one way to measure certain observables of the reduced system.

\section{Imposing Constraints}

This section contains a brief discussion of the treatment of constraints in our formalism. From the onset of our formalism, it is clear that our construction will use constraints in the full system. Before we explain the idea of applying our construction to the constraint surface, we will briefly review some notation.

\subsection{Notation}

Let us assume a set of first class classical constraints, which are smooth functions on phase space, whose Poisson action generate gauge transformations. An observable is a real function on phase space that Poisson commutes with all constraints. 
\\
There are a lot of subtleties in carrying out Diracs programme and quantizing constraints (see e.g. \cite{thiemann}) which we can not include here. We will simply assume that we are given an anomaly free set of constraint operators $\mathcal C=\{\hat C_i\}_{i \in \mathcal I}$, which are represented as operators on a dense domain in the Hilbert-space on which the quantum algebra $\mathfrak A$ is represented. Then our task is twofold:
\begin{enumerate}
  \item We want to construct the commutant $\mathcal C^\prime$ of $\mathcal C$
  $$ \mathcal C^\prime :=\{O : [C_i,O]=0 \forall i \in \mathcal I\}, $$
  whose self-adjoined elements build representatives for observables. Obviously $\mathcal C^\prime$ is a closed $*$-algebra. A physical observable is then an equivalence class of elements of the commutant $\mathcal C^\prime$, which differ only by an element of the center of the constraints.
  \item We want to find an induced Hilbert-space representation of the commutant $\mathcal C^\prime$ of the constraints, which is compatible with representation of the full theory. We will not discuss the methods used to induce such a representation, since our construction starts with a full system and assumes that we have a procedure to seek solutions to the constraints in the full system.
\end{enumerate}
Since the Schr\"odinger Hilbert space is obtained by completing the algebra of configuration variables in a positive form, we have the feeling that the induced Hilbert space representation is constructed from those configuration variables, that lie in the commutant of the constraints. However, this assertion is not always true, since the commutant $\mathcal C^\prime$ of the constraints is not always a subalgebra of $\mathfrak A$, as can be seen in may examples in which group averaging applies: e.g. by restricting quantum mechanics on $\mathbb R^2$ to quantum mechanics on $\mathbb R$, where the constraint would be $p_2\sim 0$, but there is no configuration variable $f\in C_o(\mathbb R^2)$, that commutes with this constraint.

\subsection{General Idea: Restriction to Dirac Observables}

The general idea of this section is to apply our construction of the reduced quantum theory directly to the constraint surface. Although this procedure may not always be defined, as we saw in the previous section, we will start with it and then explain its generalization:
\\
Let us suppose, that we can construct an approximate identity for the observable algebra of the full theory $id_i: i \in \mathcal I$ of the above kind i.e. $\lim_{i \in \mathcal I}\sum_{j} O_{f^i_j,f^i_j}$, where the $f^i_j$ are configuration variables that commute with the constraints. Then we are able to use the same procedure to calculate the reduced quantum algebra and induce a representation as we have done before, just restricting ourselves to configuration variables, that commute with all constraints. Specifically, for all gauge invariant configuration variables $f_1,f_2$, we construct the operators $O_{f_1,f_2}$ and we consider again the operators $O_{i(P(f_1)),i(P(f_2))}$ and canonically associate operators $T_{f_1,f_2}$ to each operator of this kind:
\begin{equation}
  \mathcal E : T_{f_1,f_2} \leftrightarrow O_{i(P(f_1)),i(P(f_2))}. 
\end{equation}
Since we are able to construct an approximate identity for the gauge invariant observables from the gauge invariant configuration variables, we can build any observable $A$, because:
\begin{equation}
  \lim_{i \in \mathcal I} id_i A = \lim_{i \in \mathcal I} \sum_j O_{f^i_j,f^i_j} A = \lim_{i \in \mathcal I} \sum_j O_{f^i_j,A^* f^i_j}.
\end{equation}
To induce a representation of the reduced theory, we can now use the same technique as above and construct an induced vacuum sate $\omega$ for a vacuum state $\Omega$ on the full theory, again by:
\begin{equation}
  \Omega(T^\eta_{f_1,f_2}):=\omega(O^\eta_{i(P(f_1)),i(P(f_2))}).
\end{equation}
With these preparations it is now easy to address the problem that occurs, when group averaging\footnote{For a brief review the method of group averaging to solve quantum constraints see \cite{marolf}.} leaves one with gauge invariant observables, that lie outside the original algebra. There are two observations that help us out:
\begin{enumerate}
  \item The result of group averaging a configuration variable yields an distributional extension of a configuration variable. This means, that we will use the image of the configuration algebra under the rigging map for our construction.
  \item Group averaging yields a vacuum state for the gauge invariant observable algebra.
\end{enumerate}
Putting these two ingredients together, we alter our prescription for a given rigging map $\eta$ to associate the operators:
\begin{equation}
  \mathcal E : T_{f_1,f_2} \leftrightarrow O_{i(P(\eta(f_1))),i(P(\eta(f_2)))}. 
\end{equation}
Moreover, we use the same formula as above to induce a gauge invariant state on the reduced algebra by simply use the same formula as above, but we use the gauge invariant vacuum state for our induction.
\\
Having a rather general construction for imposing constraints at our disposal we have to address the issue of covariance of our quantum Poisson embedding $(P,i)$ with respect to the constraints. Let us look at the classical level: if the constraints are not linear in the momenta then they generate gauge transformations, that deform the embedding of the configuration space in the full phase space. In this case it may turn out that a quantum Poisson mapping is not well defined since a gauge transformation $\tau$ changes the value of $P$, i.e. $P(f) \ne P(\tau f)$, which leads to additional spurious solutions of the constraints stemming from one and the same physical observable, creating a situation that similar to Gribov copies.
\\
A brute force solution to this problem is to construct $P$ gauge covariantly, i.e. to constrain $P(f)=P(\tau f)$ for all gauge transformations $\tau$. This ensures that the map $i$, i.e. ''assigning a representative out of the equivalence class $img(P)$'', defines a gauge covarinat quantum Poisson embedding, since $P$ acts on gauge orbits and $i$ amounts to assigning a representative to a gauge orbit. 

\subsection{Forwarding Constraints}

In this short subsection, we want to mention the possibility of ''forwarding constraints'' briefly. The idea is to use the approximate identity $id_{i \in \mathcal I}=\sum_j O_{f^i_j,f^i_j}$ and to apply the aforementioned construction of a reduced operator from a full operator to the constraints, i.e. given a constraint $C$, we associate the reduced constraint:
\begin{equation}
  C_{red}:= \lim \sum_i T_{f_i,C^* f_i}.
\end{equation}
The problem with this method is that the constraints are usually unbounded operators and thus not elements of the $C^*$-algebra. The finite gauge transformations however are supposed to be represented by bounded operators and we can apply this construction to them. Using the forwarded finite gauge transformations, we can then check, whether the reduced gauge group is continuously presented. Using these we can then find the representation of the generator and use it as the reduced constraint. Note, that this procedure works only if $(P,i)$ is a gauge covariant Poisson embedding. 

\section{Conclusion}

This work was motivated by the desire to relate Loop Quantum Cosmology \cite{bojo} with the full theory of Loop Quantum Gravity\cite{thiemann,rovelli}. Particularly, we wanted to find cosmological sectors in the theory of Loop Quantum Gravity. Since there was no prescription available that allowed for such a construction, we formulated the problem more generally.
\\
The goal of this paper was provide a construction for ''extracting relevant degrees of freedom'' from a quantum theory. We started with considering classical systems and we explained the idea that the common understanding of reducing a classical system can be understood as applying the pull-back under a Poisson embedding. This yields a reduced classical system, which is mathematically the quotient of the full observable algebra by the ideal of functions that vanish at the embedding. The physical interpretation of this situation is to say that our measurements are insensitive to this ideal.

Viewing the reduction of a classical system as the pull-back under a Poisson embedding, we identified the problem of extracting a subsystem from a quantum theory with the construction of embeddings for noncommutative spaces. We focused on transformation Lie-groupoid systems, which are a rich class of systems, that include entirely commutative spaces (i.e. the momentum group is trivial) and very noncommutative systems such as Heisenberg systems over locally compact groups. Moreover, we focused on full Poisson embeddings, which appear as the most physical embeddings, when one tries to construct subsystems of a transformation groupoid system. Moreover, we explained the canonical link between classical and quantum systems in this setting: the classical Poisson systems arise naturally as dual spaces of the associated Lie-algebroids, whereas the quantum observable algebras arise as Lie-groupoid $C^*$-algebras. This allowed us to define a classical limit as well quantization in a canonical manner.

Having the class of classical and quantum theories at our disposal, we used groupoids as mediating structures and we showed, that full transformation groupoid embeddings can be expressed on canonical groupoid modules. Using Morita theory for groupoids we where able to construct the reduced system directly from the canonical groupoid module using the re expression of the full embedding thereon. The close link between Morita theory for locally compact groupoids and Morita theory for $C^*$-algebras allowed us to read the mechanism that allows us to construct a reduced theory off from the groupoid example. For this purpose, it turned out, that a full Poisson embedding is well expressed as a pair of linear maps $(P,i)$, between the commutative algebras of configuration variables on the full and the reduced configuration space.

Using the same line of reasoning as it is done in Morita theory for $C^*$-algebras, we where then able to calculate the reduced $C^*$-algebra form the full $C^*$-algebra using the pair $(P,i)$, which we consequently identified to be a quantum Poisson embedding. Similarly, we where able to use methods analogous to Rieffel induction to induce a Hilbert space representation for the reduced $C^*$-algebra from a representation of the full $C^*$-algebra. It turned out, that the reduced algebra has the correct classical limits, i.e. that the embedded Poisson system arose as the classical limit of the reduced $C^*$-algebra. Moreover, it turned out, that the Hilbert-space representation of the reduced theory, that we constructed analogous to Rieffel induction, was such that it matched the expectation values of the full theory. The reason for all these constructions to work was the fact that we where able to use a Hilbert-$C^*$-module from the commutative algebra of functions on the quantum configuration spaces, where we where able to use the pull-back under the embedding of configuration spaces.

It turned out that our construction for the embedding of a noncommutative space amounted to a noncommutative version of the following construction for a commutative space: First construct a vector bundle over both the reduced and the full space, such that the vector bundle over the reduced space is embeddable into the vector bundle over the full space. Then construct a vector bundle embedding of the reduced vector bundle into the vector bundle over the full space. Then third recover an embedding of the reduced space into the full space by using the projection in the full vector bundle to project the embedded vector bundle down to the embedding of its base space.

An interesting twist to this discussion is given by considering transformation groupoid systems with trivial momentum group, i.e. systems based in the trivial groupoid $\mathcal G(\mathbb X)$ with $\mathcal G_o=\mathbb X$, as well as $r(x)=x,s(x)=x,e(x)=x$ and $x\circ x=x$. The $C^*$-algebra to this groupoid is $C(\mathbb X)$ and the associated canonical module is also $C(\mathbb X)$. If we now consider an embedding $\eta: \mathbb X_o \rightarrow \mathbb X$ of a space $\eta(\mathbb X_o) \subset \mathbb X$, then we obtain a pair $(P,i)$, where $P$ denotes the restriction of a function in $\mathbb X$ to $\eta(\mathbb X_o)$ and $i$ is the continuous extension of a function on $\mathbb X_o$ to a continuous function on $\mathbb X$. Now it follows immediately, that the rank-one operators $O_{i(P(f)),i(P(g))}$ for $f,g\in C(\mathbb X)$ are functions in $C(\eta(\mathbb X_o))$, given by $\overline{i(P(f))}i(P(g))$, which we can identify using $T_{f,g}$ as functions in $C(\mathbb X_o)$.

Since this work is a preparation of work on extracting cosmological sectors from Loop Quantum Gravity, we discussed the imposition of constraints in our framework. It turned out, that given a quantum Poisson embedding $(P,i)$ is covariant with respect to the constraints, we can preform our construction by simply restricting the domain of $P$ as well as the range of $i$ to the constraint surface. This is exactly the procedure, that we will use for constructing a cosmological sector in Loop Quantum Gravity in a forthcoming paper. The techniques used in this construction are similar to \cite{abl} but not equal to the ones that we use. The fundamental difference that our construction takes the full theory to the forefront and tries to embed the reduced theory explains why our construction will yield different results compared to e.g. \cite{bojo}.

\subsection*{Acknowledgements}

This work was supported by the Deutsche Forschungsgemeinschaft (DFG). I am very thankful to Martin Bojowald for inviting me to Penn State University, where part of this research was conducted and where I had the opportunity to present a preliminary version of this work, and for private discussions. I am also grateful to Thomas Thiemann for suggesting to prepare a paper that focuses on the physical ideas of this subject before presenting the mathematical results and for an invitation to the Albert-Einstein Institute in Golm near Potsdam, where he was available for discussions.

\end{document}